\documentclass[%
 aip,
 jmp,%
 amsmath,amssymb,
preprint%
]{revtex4-1}
\usepackage{graphicx}% Include figure files
\usepackage{dcolumn}% Align table columns on decimal point
\usepackage{bm}% bold math
\usepackage{subcaption}
\usepackage{setspace}
\usepackage{xcolor}
\usepackage[normalem]{ulem}
\newcommand{\etal}{{\it et al.}}

\begin{document}
%\preprint{AIP/123-QED}

%\title[Sample title]{Sample Title:\\with Forced Linebreak\footnote{Error!}}% Force line breaks with \\
%\thanks{Footnote to title of article.}
\title{Linear Stability of an Impulsively Accelerated Density Interface in an Ideal Two-Fluid Plasma}

\author{Y. Li}
% \altaffiliation{Physics Department, XYZ University.}%Lines break automatically or can be forced with \\
\affiliation{ 
Mechanical Engineering, Physical Science and Engineering Division, King Abdullah University of Science and Technology, Saudi Arabia}

\author{A. Bakhsh}
\affiliation{% 
Mathematical Sciences, Umm Al Qura University, Saudi Arabia%\\This line break forced% with \\
}
\author{R. Samtaney}%
 \email{ravi.samtaney@kaust.edu.sa.}
\affiliation{ 
Mechanical Engineering, Physical Science and Engineering Division, King Abdullah University of Science and Technology, Saudi Arabia}%

\date{\today}% It is always \today, today,

\begin{abstract}
{
We investigate the linear evolution of Richtmyer-Meshkov (RM) instability in the framework of an ideal two-fluid plasma model. The two-fluid plasma equations of motion are separated into a base state and a set of linearized equations governing the evolution of the perturbations. 
 Different coupling regimes between the charged species are distinguished based on a non-dimensional Debye length parameter $d_{D,0}$. 
When $d_{D,0}$ is large, the coupling between ions and electrons is sufficiently small that the induced Lorentz force is very weak and the two species evolve as two separate fluids. When $d_{D,0}$ is small, the coupling is strong and the induced Lorentz force is strong enough that the difference between state of ions and electrons is rapidly decreased by the force. As a consequence, the ions and electrons are tightly coupled and evolve like one fluid. 
The temporal dynamics is divided into two phases: an early phase wherein electron precursor waves are prevalent, and a post ion shock-interface interaction phase during which the RM instability manifests itself. 
We also examine the effect of an initially applied magnetic field in the streamwise direction characterized by the non-dimensional parameter $\beta_0$. For a short duration after the ion shock-interface interaction, the growth rate is similar for different initial magnetic field strengths. As time progresses the suppression of the instability due to the magnetic field is observed. The growth rate shows oscillations with a frequency that is related to the ion or electron cyclotron frequency. 
 The instability is suppressed due to the vorticity being transported away from the interface. 
}
\end{abstract}

%Use showkeys class option if keyword
\maketitle

%\singlespacing
\section{\label{sec:1}Introduction}
%Inertial confinement fusion (ICF) is a promising method for the generation of fusion energy. By imploding the target to very high densities, it is expected that the fusion reactions occur and the fuel is confined by its own inertia. A key bottleneck towards the achieving goal of ICF is hydrodynamic instabilities, such as Richtmyer-Meshkov (RM) instability which occurs when a perturbed density interface is impulsively accelerated {\citep{richtmyer1960, meshkov1969}}. {\color{red}{Numerous researches have been into the subject in the framework of hydrodynamics {\citep{guan2020, ding2021}}.}} 
{{The Richtmyer-Meshkov (RM) instability occurs when a perturbed density interface is impulsively accelerated and has been numerously investigated {\citep{richtmyer1960, meshkov1969, yang1994, zabusky1999, guan2020, ding2021}}, since it is a key bottleneck towards the successful ignition of inertial confinement fusion (ICF). ICF is a promising method for the generation of fusion energy. By imploding the target to very high densities, it is expected that the fusion reactions occur and the fuel is confined by its own inertia. Due to the high temperature and high energy-density scenario in ICF, it is expected the materials to be in a plasma state, and thus could be influenced by a magnetic field. }
}
An effective fluid description for the plasma is single-fluid magnetohydrodynamics (MHD). In the context of ideal MHD, it was demonstrated that the magnetic field suppresses the RM instability {\citep{samtaney2003}}. This was followed by a linear analytical model for RM instability in incompressible MHD {\citep{wheatley2005}}, and other studies {\citep{wheatley2009, qin2021, zhang2020}}. The single-fluid MHD studies concluded that the essential physical mechanism suppressing the instability  in the presence of  a magnetic field is due to the transport of the baroclinically generated vorticity away from the interface by MHD waves. A purely numerical approach for linear analysis of the RM instability was developed for hydrodynamics and MHD {\citep{samtaney2009}}. This numerical approach was then used to investigate RM and Rayleigh-Taylor instabilities in cylindrical geometry by
Bakhsh \etal \citep{bakhsh2016} and Baksh \& Samtaney \citep{bakhsh2018}. % performed a linear analysis of the RM instability in the context of single-fluid MHD in cylindrical geometry. 

The aforementioned MHD investigations do not take into account the effect of finite Larmor radius. In fact, the single-fluid MHD model is valid when the plasma length scales, such as Debye length and Larmor radius, are negligible compared to the characteristic length scale of the flow.  Magnetized
implosion experiments have demonstrated that the Larmor radius of alpha particles may be
larger than the hot spot size \citep{hohenberger2012}, suggesting that single-fluid MHD may not be sufficient to model the physics under this circumstance. Moreover, unless the Biermann battery effect is included \citep{srinivasan2012}, MHD fails to capture the phenomenon of self-generated electromagnetic fields. To consider the effect of plasma length scales, the two-fluid plasma model is employed. In this model, ions and electrons are treated as two separate fluids and are coupled to the full Maxwell equations. In addition, the electron particle mass and light speed are finite. The two-fluid plasma model also allows for the investigation of self-generated electromagnetic fields. 
Bond \etal\cite{bond2017} investigated the multi-fluid plasma RM instability of a thermal interface and noted that the two-fluid plasma differed  significantly from the hydrodynamic case.  Nonlinear simulations by Bond \etal~\cite{bond2020} showed that the two-fluid plasma RM instability is suppressed by an initially imposed magnetic field with increasing effectiveness as plasma length scale is decreased. 

Presently, the linear stability of an impulsively accelerated density interface is investigated in the framework of the ideal two-fluid plasma. 
{{Although the existing nonlinear stability precedes the linear work, and since it is obvious that nonlinear effects do not manifest themselves in linear studies, it does not imply linear analysis is not necessary. Two fluid plasma nonlinear simulations show that the interface dynamics is extremely complex. In the spirit of a reductionist investigation, linear stability simulations shed light on the instability and its suppression by reducing the complexity encountered in nonlinear simulations and hence better elucidate physical mechanisms.
In addition, linear simulations enable us to explore a wider range of parameters compared with earlier nonlinear simulations due to the relatively inexpensive computational cost. Furthermore, the spatial and temporal resolution in linear studies can vastly exceed those employed in nonlinear simulations. Linear studies, such as the present one, can serve as a guide for future nonlinear simulations. In addition, the work on linear stability fills an existing knowledge gap in the sense that there is no prior work on linear stability of interfaces with the two fluid plasma model.
%Linear simulations enable us to explore a wider range of parameters compared with earlier nonlinear simulations due to the relatively inexpensive computational cost. Moreover, the spatial and temporal resolution in linear studies can vastly exceed those employed in nonlinear simulations. Linear studies, such as the present one, can serve as a guide for future nonlinear simulations. In addition, due to the complex physics encountered in two-fluid plasma RM instability investigations, linear studies can help better elucidate physical mechanisms since the complexity of the physics is somewhat decreased in the context of linear studies. 
In the present work, we enhance the numerical method developed by Samtaney \citep{samtaney2009} to take into account the expanded set of linearized two-fluid plasma equations. A brief comparison between linear and nonlinear simulations is discussed in Appendix \ref{apdx_comparison}.
}}

The remainder of the paper is organized as follows: The original and linearized ideal two-fluid plasma models, numerical method and initial setup are introduced in Section \ref{sec:2}. In Section \ref{sec:3} we present linear simulation results in the absence of an initial magnetic field and discuss the different coupling regimes by varying the  reference Debye length. In \ref{sec:4}, linear simulation results are presented when there is an initial magnetic field present and the growth rate of the perturbations is examined for different initial magnetic field strength for weak and strong coupling between the ions and electrons.  Conclusions are presented in Section \ref{sec:5}.
%%%%%%%%%%%%%%%%%%%%%%%%%%%%%%%%%%%%%%%%%%%%%%%%%%%%%%%%%%%%%%%%%%%%%%
\section{\label{sec:2}Linearization of Two-fluid Plasma Equations and Numerical Details}

\subsection{\label{sec:2.1}Two-fluid plasma model}
We use an ideal two-fluid plasma model for this work. In this model, the collisional equilibrium state is instantaneously reached in each species while no collisions are considered between particles of different species. Therefore, the ions and electrons are treated as two separate fluids described by Euler equations with the Lorentz force as the source term,
\begin{eqnarray}
&&\frac{\partial \rho_\alpha}{\partial t}+\nabla \cdot \left(\rho_\alpha \bm{u}_\alpha \right) = 0, \\
&&\frac{\partial \rho_\alpha \bm{u}_\alpha}{\partial t}+\nabla \cdot \left(\rho_\alpha \bm{u}_\alpha \bm{u}_\alpha +p_\alpha \bm{I} \right) = n_\alpha q_\alpha \left(\bm{E}+\bm{u}_\alpha \times \bm{B} \right),\\
&&\frac{\partial \mathcal{E}_\alpha}{\partial t}+\nabla \cdot \left(\left({\mathcal{E}_\alpha} + p_\alpha \right)\bm{u}_\alpha\right) = n_\alpha q_\alpha \bm{E} \cdot \bm{u}_\alpha,
\end{eqnarray}
where,
\begin{equation}
\rho_\alpha=n_\alpha m_\alpha, \quad p_\alpha=n_\alpha k_B T_\alpha, \quad {\mathcal{E}_\alpha}=\frac{p_\alpha}{\gamma_\alpha-1}+\frac{\rho_\alpha |\bm{u}_\alpha|^2}{2}.
\end{equation}
The subscript $\alpha$ denotes the species with `$\alpha=i (e)$' for ions (electrons). $\rho$, $n$, $m$, $\bm{u}=\left(u, v, w \right)^T$, $p$, $\mathcal{E}$, $q$ and $T$ are the density, number density, particle mass, velocity, pressure, energy, particle charge and temperature, respectively. $\gamma$ is specific heat ratio with the value of $5/3$ for each species throughout this study. $k_B$ is the Boltzmann constant. Since inter-species collisions are not considered, the interactions between ions and electrons are via the induced magnetic field $\bm{B}$ and electric field $\bm{E}$. The evolution of electromagnetic field is governed by the Maxwell equations with two correction potentials $\psi_B$ and $\psi_E$ for divergence constraints \citep{munz2000}.
\begin{eqnarray}
&&\frac{\partial \bm{B}}{\partial t}+\nabla \times \bm{E} +\Gamma_B \nabla \psi_B= \bm{0}, \\
&&\frac{\partial \bm{E}}{\partial t}- c^2 \nabla \times \bm{B} +c^2\Gamma_E \nabla \psi_E= -\frac{1}{\epsilon_0} \sum_{\alpha}n_\alpha q_\alpha \bm{u}_\alpha, \\
&&\frac{\partial \psi_E}{\partial t}+\Gamma_E \nabla \cdot \bm{E} = \frac{\Gamma_E}{\epsilon_0}\sum_{\alpha}n_\alpha q_\alpha, \\
&&\frac{\partial \psi_B}{\partial t}+c^2 \Gamma_B\nabla \cdot \bm{B} = 0,
\end{eqnarray}
where $c=1/\sqrt{\mu_0 \epsilon_0}$ is the light speed with $\mu_0$  permeability of free space and $\epsilon_0$ vacuum permittivity. The introduced correction potentials $\psi_B$ and $\psi_E$ serve to enforce the divergence constraints by transferring the divergence errors out of the domain with the speed $\Gamma_B c$ and $\Gamma_E c$, respectively. Here, $\Gamma_B $ and $\Gamma_E $ are chosen to be unity throughout this study. 

The dimensionless variables are defined as below by specifying the reference variables (with subscript $0$), :
\begin{eqnarray}
&&\hat{\bm{x}}=\frac{\bm{x}}{L_0},~ \hat{t}=\frac{t}{L_0/u_0},~ \hat{\rho}_\alpha=\frac{\rho_\alpha}{n_0m_0},~ \hat{m}_\alpha=\frac{m_\alpha}{m_0},~ \hat{\bm{u}}_\alpha=\frac{\bm{u}_\alpha}{u_0},~ \hat{q}_\alpha=\frac{q_\alpha}{q_0},~ \hat{p}=\frac{p_\alpha}{n_0 m_0 u_0^2},~\nonumber \\
&&\hat{\bm{B}}=\frac{\bm{B}}{B_0},~ \hat{\bm{E}}=\frac{\bm{E}}{c B_0},~ \hat{\psi}_E=\frac{\psi_E}{B_0},~ \hat{\psi}_B=\frac{\psi_B}{c B_0},~ \hat{c}=\frac{c}{u_0},
\end{eqnarray}
where the reference magnetic field $B_0=\sqrt{\mu_0 n_0 m_0 u_0^2}$. Therefore, the dimensionless ideal two-fluid plasma equations with the above notation may be written as follows, with the carets omitted for simplicity.
\begin{eqnarray}
\label{eq:density}
&&\frac{\partial \rho_\alpha}{\partial t}+\nabla \cdot \left(\rho_\alpha \bm{u}_\alpha \right) = 0, \\
\label{eq:momentum}
&&\frac{\partial \rho_\alpha \bm{u}_\alpha}{\partial t}+\nabla \cdot \left(\rho_\alpha \bm{u}_\alpha \bm{u}_\alpha +p_\alpha \bm{I} \right)= \frac{n_\alpha q_\alpha}{d_{D,0} c} \left(c\bm{E}+\bm{u}_\alpha \times \bm{B} \right),\\
\label{eq:energy}
&&\frac{\partial \mathcal{E}_\alpha}{\partial t}+\nabla \cdot \left(\left({\mathcal{E}_\alpha} + p_\alpha \right)\bm{u}_\alpha\right) =  \frac{n_\alpha q_\alpha }{d_{D,0}} \bm{E} \cdot \bm{u}_\alpha,\\
\label{eq:magnetic}
&&\frac{\partial \bm{B}}{\partial t}+c\nabla \times \bm{E} +c\Gamma_B \nabla \psi_B= \bm{0}, \\
\label{eq:electric}
&&\frac{\partial \bm{E}}{\partial t}- c \nabla \times \bm{B} +c\Gamma_E \nabla \psi_E= -\frac{1}{d_{D,0}} \sum_{\alpha}n_\alpha q_\alpha \bm{u}_\alpha, \\
\label{eq:psi_E}
&&\frac{\partial \psi_E}{\partial t}+c\Gamma_E \nabla \cdot \bm{E} = \frac{c \Gamma_E}{d_{D,0}}\sum_{\alpha}n_\alpha q_\alpha, \\
\label{eq:psi_B}
&&\frac{\partial \psi_B}{\partial t}+c \Gamma_B\nabla \cdot \bm{B} = 0.
\end{eqnarray}
Two dimensionless parameters arise from the non-dimensional process. One of these, that appears in the equations above, is the reference Debye length $d_{D,0}=\sqrt{\frac{\epsilon_0 m_0 u_0^2}{n_0 q_0^2 L_0^2}}$. Noting that the electromagnetic force is essentially inversely proportion to $d_{D,0}$, the magnitude of $d_{D,0}$ dictates the coupling between ions and electrons to some extent. A large Debye length implies weak coupling between the two charged species. As the $d_{D,0}\rightarrow\infty$, the coupling decreases to zero, as if the two charged fluids do not interact  with the electromagnetic fields, as seen in Eqs. (\ref{eq:momentum}) and (\ref{eq:energy}) where the source terms approach $0$. In this limit, ions and electrons evolves like two uncoupled hydrodynamic fluids. On the other hand, the limiting behavior of $d_{D,0}\rightarrow 0$ is that the coupling is so strong that the ions and electrons essentially evolve together as a ``single" fluid. The second dimensionless parameter , which does not appear in the equations,  is 
$\beta_0=\frac{2(p_i+p_e)}{|\bm{B}|^2}$. This is related to the initial strength of the applied magnetic field. When $\beta_0$ is infinity, it implies that no initial magnetic field is presented in simulations. We note here that one may choose a reference $B_0$ scale independently in which case $\beta_0$ will appear in the equations. 

\subsection{\label{sec:2.2}Linearization}
The non-dimensional equations in conservative form in two dimensions can be written as follows,
\begin{equation}
\label{eq:mainEquation}
\frac{\partial {U}}{\partial t} + \frac{\partial {F}({U})}{\partial x} + \frac{\partial {G}({U})}{\partial y} = {S}. 
\end{equation}
We linearize the ideal two-fluids plasma equations by splitting the solution vector into a base and perturbed solutions, i.e.  ${U}={U}^\circ+\epsilon \hat{{U}}\exp(iky)$. Here, $k$ is the wave number and $\epsilon$ is a small value ($\epsilon$ may be related to the initial small amplitude  of the density perturbed interface). We note that the base state is time-dependent. Substituting it into {Eq.} (\ref{eq:mainEquation}), we derive{ a set }of nonlinear equations for the base state and another set of linear equations governing the perturbations, respectively. 
\begin{subequations}
\begin{equation}
\label{eq:base}
\frac{\partial {U}^\circ}{\partial t}+\frac{\partial{F}({U}^\circ) }{\partial x}={S}({U}^\circ),
\end{equation}
\begin{equation}
\label{eq:perturb}
\frac{\partial \hat{{U}}}{\partial t}+\frac{\partial{A}({U}^\circ) \hat{{U}}}{\partial x}=(-ik{B}({U}^\circ)+{C}({U}^\circ))\hat{{U}},
\end{equation}
\end{subequations}
where ${A}=\frac{\partial{F}}{\partial U}|_{{U}^\circ }, {B}=\frac{\partial{G}}{\partial U}|_{{U}^\circ }$ and ${C}=\frac{\partial{S}}{\partial U}|_{{U}^\circ }$ are the Jacobean matrices. The details of $A$, $B$, and $C$ are given in Appendix \ref{apdx_matrix}. Here we note that the source term ${C}({U}^\circ))\hat{{U}}$ plays an important role in the dynamics of the perturbed quantities: this source term matrix ${C}({U}^\circ))$ is essentially the forcing on the perturbations due to the base state Lorentz force. 

Due to the finite speed of light and the existence of light waves in the two-fluid plasma model, we find that the light and fast electron waves tend to reflect off the domain boundary, causing unphysical oscillations. One method to mitigate this is to develop better absorbing boundary conditions while the other is to use an unbounded domain in $x$. We choose the latter approach and map $x\in(-\infty,\infty)$ to $\xi\in(-1,1)$, where the mapping variable is $\xi(x)=\frac{2}{\pi}\arctan(\sigma x)$ ($\sigma = 0.1$ is chosen for the simulations presented later). 
%During the simulations, the light wave and fast electron waves may reflect from the boundaries and lead to the unphysical oscillations in the results. The stretched mesh $\xi(x)=\frac{2}{\pi}\arctan(\sigma x)$ with $\sigma = 0.1$ is then used. By mapping $x\in(-\infty,\infty)$ to $\xi\in(-1,1)$, the influences of the boundaries are eliminated. 
Then the Eqs. (\ref{eq:base}) and (\ref{eq:perturb}) become
\begin{subequations}
\begin{equation}
\label{eq:streqbase}
\frac{\partial x_\xi {U}^\circ}{\partial t}+\frac{\partial{F}({U}^\circ) }{\partial x}=x_\xi {S}({U}^\circ),
\end{equation}
\begin{equation}
\label{eq:streqpert}
\frac{\partial x_\xi \hat{{U}}}{\partial t}+\frac{\partial{A}({U}^\circ) \hat{{U}}}{\partial x}=x_\xi(-ik{B}({U}^\circ)+{C}({U}^\circ))\hat{{U}}.
\end{equation}
\end{subequations}

\subsection{\label{sec:2.3}Physical setup}
\begin{figure}
\includegraphics[width=\linewidth]{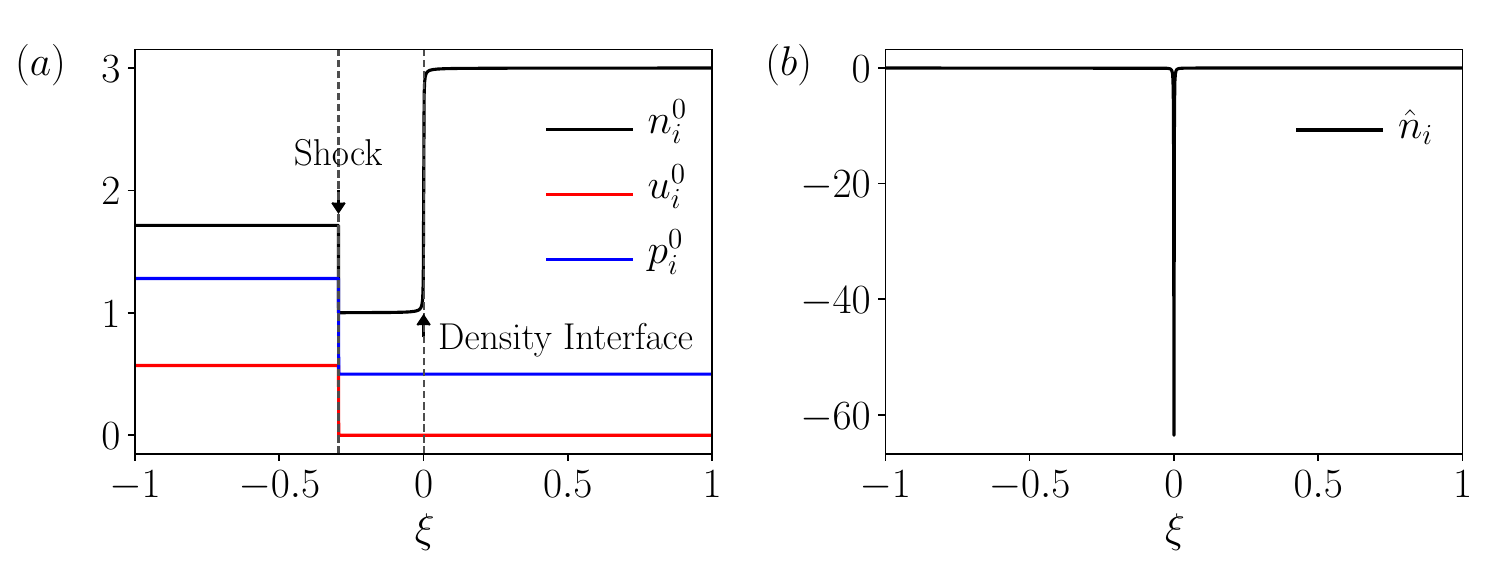}
\caption{\label{fig:setup} Initial setup of the ions with ion shock strength $M_s=1.5$;  (a) base state of ion , (b) perturbation of number density. The figures are plotted in stretched coordinates.}
\end{figure}

In this study, we consider the shock-interface interaction cases in a two-fluid plasma. As shown in Fig. \ref{fig:setup}, the whole domain is divided into three sections by an ion shock and a density interface. An ion shock with strength of $M_s=1.5$ initialized at $\xi=-0.295$ $(x\approx-5.0)$ moves from left to right, and interacts with the ion density interface centered at $\xi=0$ $(x=0)$. For a sharp density interface, the perturbed number density profile across the ion interface can be written with a Heaviside function as follows,
\begin{equation}
\label{eq:denfile}
n_i(x(\xi),y) = \frac{n_{i,1}+n_{i,2}}{2}+\frac{n_{i,2}-n_{i,1}}{2}H(x(\xi)-\epsilon \exp(iky)),
\end{equation}
where $n_{i,1}$ and $n_{i,2}$ are the unshocked number density to the left and right of the interface of ions with the values set to be $1.0$ and $3.0$, respectively. Here, $\epsilon$ denotes the perturbation amplitude. Note that in linear stability, the perturbation amplitude $\epsilon$ cancels out in the linear set of equations governing the perturbations (\ref{eq:perturb}). In the results presented later in Sections \ref{sec:3} and \ref{sec:4}, we will plot the growth rate of the amplitude and the time history of the amplitude normalized by its initial value: these are referred to as the ``perturbation growth rate" and ``perturbation amplitude". %For the sake of clarity, we distinguish these two quantities from the perturbation states (these are the quantities with the ``$\hat{}$". 
Instead of the strictly sharp interface, we use a smooth approximation to the step function $H(x(\xi))$. This so-called regularized Heaviside function is then defined as $H(x(\xi)) = \frac{2}{\pi} \arctan(\frac{x(\xi)}{\delta})$, where $\delta$ is a measure of the smoothness applied to the sharp interface. 
As a result, by retaining only the linear term in a Taylor expansion in Eq. (\ref{eq:denfile}), we have the initial number density profile of the base state across the interface as,
\begin{equation}
\label{eq:denbase}
n_i^0(x(\xi)) = \frac{n_{i,1}+n_{i,2}}{2}+\frac{n_{i,2}-n_{i,1}}{\pi}\arctan(\frac{x(\xi)}{\delta}),
\end{equation}
and the perturbation of number density $\hat{n}_i$ as,
\begin{equation}
\label{eq:denpert}
\hat{n}_i(x(\xi)) = \frac{n_{i,1}-n_{i,2}}{\pi}\frac{\delta}{x(\xi)^2+\delta^2}.
\end{equation}
Presently we choose $\delta = 0.01$ throughout this work. Ahead of the ion shock, we set the ion pressure $p_{i,1}=p_{i,2}=0.5$ and velocity $\bm{u}_{i,1}=\bm{u}_{i,2}=\bm{0}$. Therefore, the ion states behind the shock are,
\begin{equation}
\label{eq:state}
n_{i,0}= R ~n_{i,1}; \quad \bm{u}_{i,0}=(\dfrac{R-1}{R}M_s \sqrt{\dfrac{\gamma_i p_{i,1}}{\rho_{i,1}}},0,0)^T;\quad p_{i,0} = \dfrac{(\gamma_i +1)R-(\gamma_i-1)}{(\gamma_i+1)-(\gamma_i-1)R}~p_{i,1};
\end{equation}
with $R=\frac{(\gamma_i+1)M_s^2}{2+(\gamma_i-1)M_s^2}$. Initially,  charge neutrality, thermal equilibrium and mechanical equilibrium are satisfied in each section, viz., $n_{i,s}=n_{e,s}$, $p_{i,s}=p_{e,s}$ and $\bm{u}_{i,s}=\bm{u}_{e,s}$, $s=0,1,2$. For investigations of an initially imposed magnetic field, the initial magnetic field is applied along the $x$ direction, with the magnitude determined as $B_x=\sqrt{\frac{2(p_i+p_e)}{\beta_0}}$, where $p_i$ and $p_e$ are the pressures on the density interfaces. 

The non-dimensional electron charge $q_e$ and ion charge $q_i$ are $-1$ and $1$, respectively. We use the mass ratio $m_i/m_e=100$ instead of the physical value $1836$, ($m_i$ is normalized to unity) to reduce the problem stiffness while ensuring relatively fast electron dynamics \citep{bond2017, yuan2020, bond2020}. Based on the limiting values of hotspot temperature and number density in ICF implosion ($T_0=5 \times 10^3$ eV with $n_0=10^{31}$) \citep{srinivasan2012}, we set the non-dimensional light speed $c=50$ to reduce the computational cost. By varying the reference Debye length $d_{D,0}$, we investigate the two-fluid effect on the perturbation growth of the RM instability. Moreover, the effect of the initial applied magnetic field on the growth of the perturbations is examined by varying $\beta_0$.
For simplicity, the wave number of the perturbation is fixed as $k=2\pi$ for all the simulations.

\subsection{\label{sec:2.4}Numerical Implementation}
A third-order TVD Runge-Kutta \citep{gottlieb1998}  scheme is used to solve the Eqs. (\ref{eq:streqbase}) and (\ref{eq:streqpert}), with HLLC \citep{toro1994} solver for the fluid fluxes and HLLE \citep{einfeldt1988} scheme for the electromagnetic fluxes, while the Roe solver is applied for the perturbation fluxes. The source terms are treated locally with an implicit method \citep{abgrall2014}. For all the simulations presented, during the entire time duration in each simulation,  the effective resolution is at least {$4000$} cells per unit length {{(see Appendix \ref{apdx_conv} for convergence test)}}. A volume-of-fluid approach is used to track the density interface in each species, where the tracer variable $\phi_\alpha \in\left[-1,1\right]$. %For simplicity, we map the stretched coordinate back to the normal one in the following discussions. 

%\section{\label{sec:3}Results and Discussions}
\section{\label{sec:3}Zero Initial Magnetic Field \boldmath{$\beta_0=\infty$}}

\begin{figure}
\includegraphics[width=\linewidth]{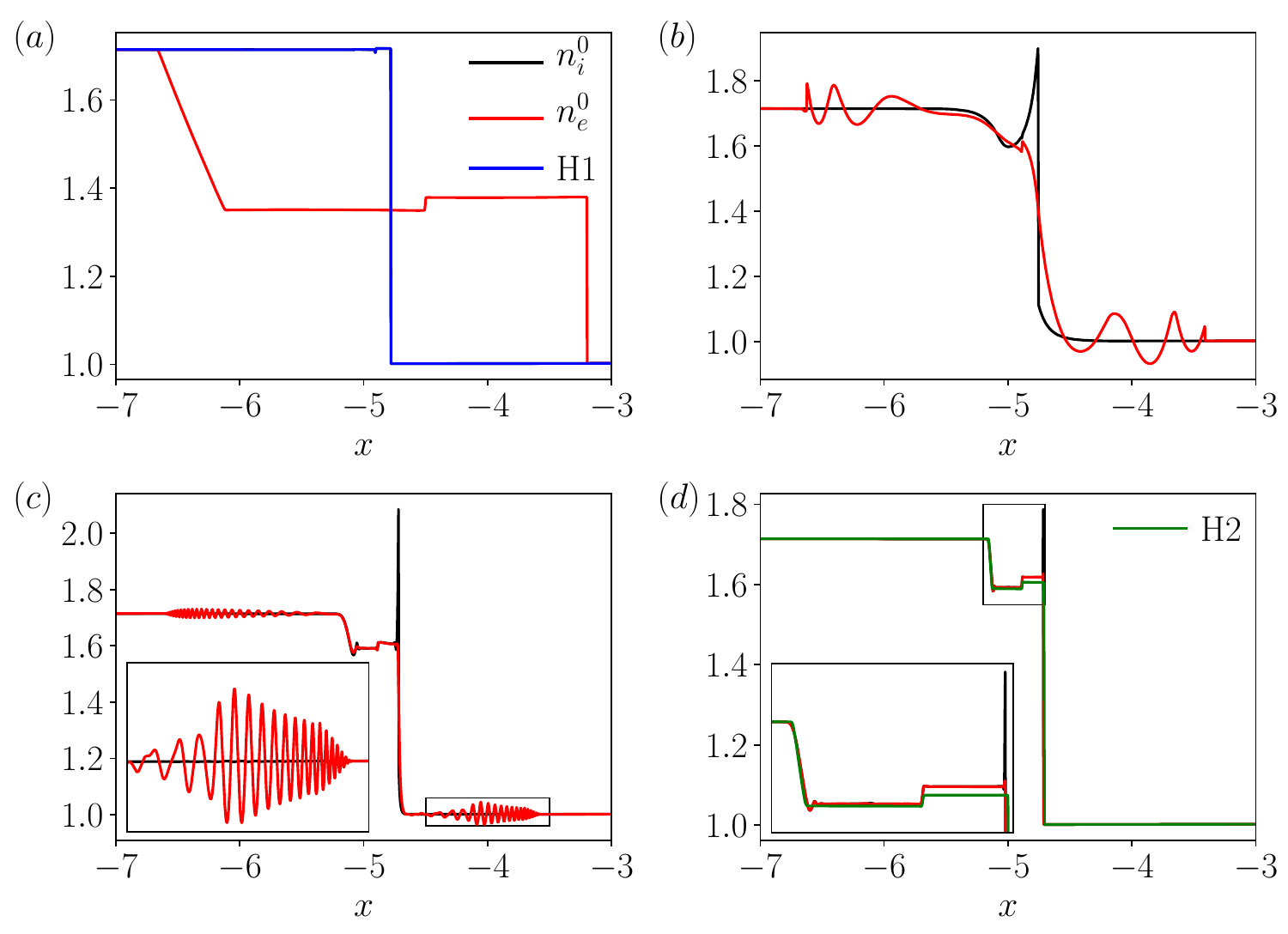}
\caption{\label{fig:numdenbase} Base number density of ions and electrons at $t=0.157$ for the cases with various reference Debye lengths; (a) $d_{D,0}=10$, (b) $d_{D,0}=0.1$, (c) $d_{D,0}=0.01$, (d) $d_{D,0}=0.001$. The results are compared with those of limiting hydrodynamic cases `H1' and `H2'. }
\end{figure}

\begin{figure}
\includegraphics[width=\linewidth]{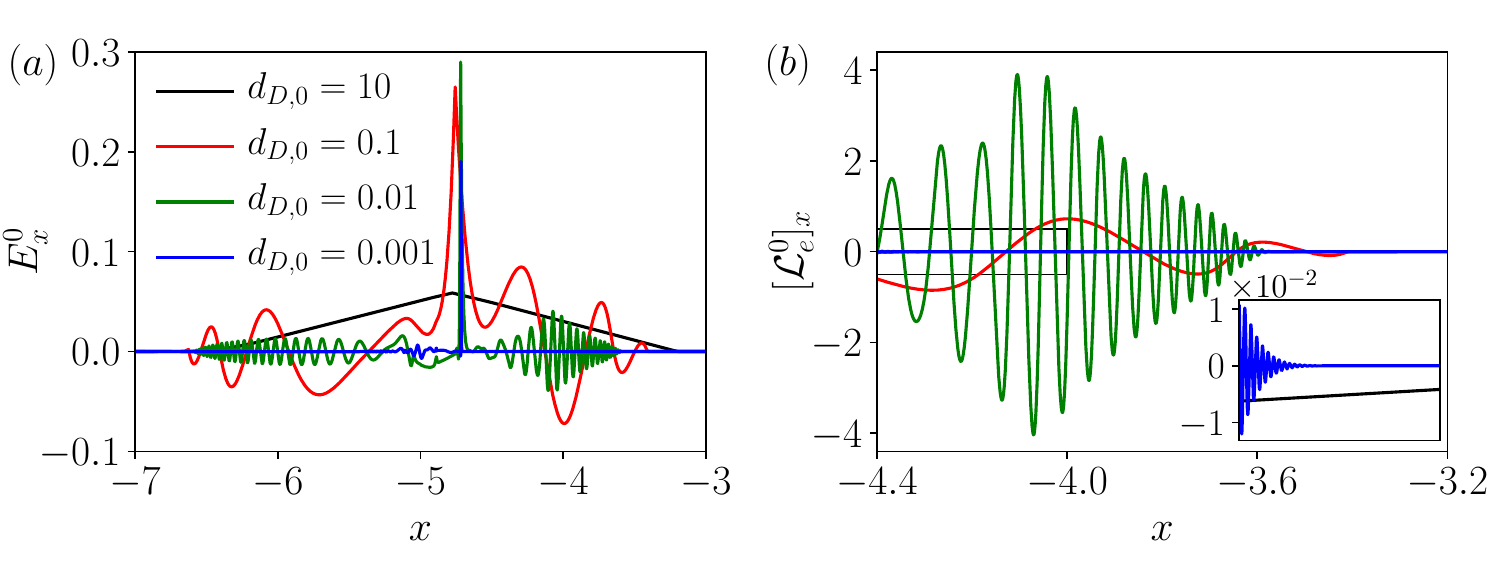}
\caption{\label{fig:lorfbase} Base electric field and electron Lorentz force along $x$ direction at $t=0.157$ for the cases with various reference Debye lengths: (a) $E_x^0$, (b) $[\mathcal{L}_e^0]_x$. }
\end{figure}
In this section, we present results for the cases in which no initial magnetic field is applied, i.e. $\beta_0=\infty$. 
We first present a discussion of the various coupling regimes between the ion and electron species as characterized by the parameter $d_{D,0}$. This is followed by examining the weakly coupled ($d_{D,0}=0.1$)  and strongly coupled cases ($d_{D,0}=0.01$) in more detail. In each of these cases, we distinguish between two temporal phases. 
The first, and early phase is associated with the interactions of electron waves with the electron interface. This is dubbed the ``electron precursor waves'' phase. The second phase, which is more dynamically important for the growth of the ion interface perturbations, is associated with the time period after the interaction of the ion shock with the ion density interface, and is dubbed the ``post-ion shock interaction" phase. Furthermore, for each of the weakly and strongly coupled cases, we examine the temporal history of the  perturbation growth rate and the normalized amplitude. We correlate the oscillations in spatial profiles and temporal histories with different frequencies that are present in the two-fluid plasma system. 

\subsection{Coupling Regimes}
The different coupling regimes are illustrated by examining the wave structures in each of the ion and electron fluids.  
We examine the base state prior to the ion shock-interface interactions, and plot the spatial profile of density for ions and electrons at $t=0.157$ for the cases with various $d_{D,0}$ in Fig. \ref{fig:numdenbase}. For comparison, two asymptotic hydrodynamic limits of the two-fluid plasma model are considered. For $d_{D,0} \rightarrow \infty$,  other parameters remaining finite, the ions and electrons are fully decoupled and evolve as two separate fluids which  are not influenced  by electromagnetic forces. In this limit, we refer to the ion part as the case ``H1", in which a single shock with $M_s=1.5$ travels towards the density interface (plotted in Fig. \ref{fig:numdenbase}(a)). 
The other extreme limit arises as the reference light speed $c \rightarrow \infty$ and $d_{D,0}c \rightarrow 0$: in this limit the ideal two-fluid plasma model degenerates to the hydrodynamic equations, and the initial density, momentum and pressure in each section becomes $\sum_\alpha \rho_\alpha$, $\sum_\alpha \rho_\alpha \bm{u}_\alpha$ and $\sum_\alpha p_\alpha$, respectively \citep{shen2018}. This limiting case is referred to as ``H2". In this case, a shock, contact discontinuity and rarefaction wave are generated from the Riemann interface, see Fig. \ref{fig:numdenbase}(d).
For large values of the parameter, say $d_{D,0}=10$, the coupling between ions and electrons is sufficiently weak that the two species barely influence each other. In this large   $d_{D,0}$ case, as shown in Fig. \ref{fig:numdenbase}(a),  the shock in the ion fluid has translated a small distance and the ion shock profile and position matches well with that of ``H1" case. In contrast, the much lighter electrons have evolved signficantly.  The distinct shock, contact discontinuity and rarefaction wave in electron wave structure is shown in Fig. \ref{fig:numdenbase}(a). 

As $d_{D,0}$ decreases, the extent of coupling between the two species increases. In particular, the electromagnetic forces influence both species, with the effects on the electrons much more prominent due to their smaller particle mass. To demonstrate this coupling we consider $d_{D,0}=0.1$. In contrast to the $d_{D,0}=10$ case, ions and electrons significantly impact each other through the induced electromagnetic force in this case,  resulting in a significant transformation of the wave structures. The ion and electron number density for this lower $d_{D,0}$ case is plotted in Fig. \ref{fig:numdenbase}(b). 
%The electron waves appeared in the $d_{D,0}=0$ case disappear and the ions and electrons show an obviously coupling phenomenon, see Fig. \ref{fig:numdenbase}(b). 
It shows that the electron number density oscillates about the ion density. We note that although the number density of ions seems to be constant in the region where electrons number density oscillates, it actually oscillates but with much smaller amplitude due to the $100$ times heavier ion particle mass. Decreasing $d_{D,0}$ further to $0.01$, we observe that the wavenumber of the electrons waves increases while the amplitude decreases in general, and the configuration of electron wave structures is almost same as ions (Fig. \ref{fig:numdenbase}(c)). Finally, in the case with $d_{D,0}=0.001$, the coupling effect is sufficiently strong enough that the number densities of ions and electrons matches well except in the very narrow region right behind the ion shock (see inset in Fig. \ref{fig:numdenbase}(d)). However, the profiles $d_{D,0}=0.001$ are not completely congruent with the ``H2" profiles because of the insufficiently large light speed ($c=50$). 

It is instructive to examine the induced electromagnetic forces between ions and electrons for a deeper understanding of the behavior of the two species in Fig. \ref{fig:numdenbase}.
An examination of the two-fluid plasma equations leads us to the conclusion that no electromagnetic fields except the $x-$ direction aligned electric field will be induced in the base state when $\beta_0=\infty$. Figure \ref{fig:lorfbase} plots the electric field and electron Lorentz force along the $x$ direction in the base state at $t=0.157$ for various values of the coupling parameter $d_{D,0}$. According to Eq. (\ref{eq:electric}), the growth of $E^0_x$ is proportional to the $x- $ component of the current, $\sum_\alpha n_\alpha q_\alpha u_\alpha$, (denoted as $j^0_x$ henceforth) while it is inversely proportional to $d_{D,0}$. The difference in this source term between the ions and electrons is at its maximum at the location of the ion shock, and $j^0_x$ is the maximum there. Therefore, $E^0_x$ has its peak value at the ion shock (see Fig. \ref{fig:lorfbase}(a)). Although the value of $j^0_x$ at the ion shock is larger for the case with smaller $d_{D,0}$, the stronger coupling effect decreases the value of $j^0_x$ in a shorter time duration that the peak of $j^0_x$ ($E^0_x$) at the ion shock is of the same order of magnitude for all four different values of  $d_{D,0}$, as shown in Fig. \ref{fig:lorfbase}(a). The induced electric field $E^0_x$ leads to the electromagnetic force exerted on ions and electrons. According to  Eq. (\ref{eq:momentum}), the corresponding Lorentz force $[\mathcal{L}^0_\alpha]_x$ is $\frac{n_\alpha^0 q_\alpha E^0_x}{d_{D,0}}$, which is inversely proportional to $d_{D,0}$. Figure \ref{fig:lorfbase}(b) shows the electron Lorentz force in the region away from the ion shock for the cases with different $d_{D,0}$. For the case with $d_{D,0}=10$, though the induced electric field $E^0_x$ and number density $n_e$ is comparable to those of case with $d_{D,0}=0.1$, the Lorentz force $[\mathcal{L}^0_e]_x$ is smaller than $1\times10^{-2}$ due to the much larger reference Debye length. Hence, the Lorentz force has an insignificant influence on the evolution of electrons, not to mention ions. As $d_{D,0}$ decreases, the influence of Lorentz force on the charged species increases, and 
the difference between the states of two species is reduced due to the $[\mathcal{L}^0_e]_x$ that is induced by the electric field $E_x^0$.
 However, due to the inertia of the electrons, the difference not only continues to decrease but eventually goes through a sign change and increases in magnitude (the sign of $j^0_x$ and $[\mathcal{L}^0_\alpha]_x$ reverses). This results in an overshoot, so that $E_x^0$ gradually decreases to $0$ and its magnitude increases albeit with a sign reversal. Finally, the reversed $-E^0_x$ induces a Lorentz force that reverses the process discussed above. Essentially, in  one whole cycle,  the above process is $-j^0_x\rightarrow E^0_x\rightarrow -[\mathcal{L}^0_e]_x \rightarrow j^0_x \rightarrow -E^0_x \rightarrow [\mathcal{L}^0_e]_x \rightarrow -j^0_x $. It means that the electron state and induced electric field oscillate over time which is manifested as the oscillations in the spatial profiles. Therefore, we can see the oscillation waves in electron number density and electric field in Figs. \ref{fig:numdenbase} and \ref{fig:lorfbase}. It is noted that the above physical process results also in oscillations of the ion states, however, due to the $100$ times larger particle mass, the oscillation amplitude of ions is negligible compared with the electrons. As a result, the ion number density $n_i$, looks somewhat ``flat" compared with the $n_e$ in Fig. \ref{fig:numdenbase}. We may estimate the frequency $\Omega_E$ and wave length $\lambda_E$ of the oscillation waves in $E^0_x$ by combining Eqs. (\ref{eq:momentum}) and (\ref{eq:electric}) while neglecting the gradient terms,
\begin{equation}
\label{eq:Eosci}
\frac{\partial^2 E_x^0}{\partial t^2}+\dfrac{1}{d_{D,0}^2}(\frac{n_i}{m_i}+\frac{n_e}{m_e})E^0_x=0.
\end{equation}
Thus the frequency of $E_x^0$ waves is $\Omega_E= \frac{\sqrt{n_i+\frac{m_i}{m_e} n_e}}{d_{D,0}}$ and the wave length $\lambda_E$ is $\approx \frac{2\pi c_{s,e}}{\Omega_E}$, where $c_{s,e}$ is the sound speed in electrons. Since the magnitudes of $n_i$, $n_e$ and $u_e$ are of same orders respectively for all cases, we can draw an approximate conclusion that the frequency and wave length of the oscillation waves are inversely proportional to the reference Debye length $d_{D,0}$, which is confirmed to some extent in the Fig. \ref{fig:lorfbase}(b).  

Although the electrons oscillate under the Lorentz force, the long term effect of $[\mathcal{L}^0_e]_x$ is to bind the two species. Therefore, as time progresses, the difference between ions and electrons gradually shrinks to $0$, brings about the gradually decreasing electric field that results in the gradually damping oscillations of $n_e$. The smaller $d_{D,0}$ is, the sooner is the combination of the two species and  shorter the duration  during which the oscillations get damped. As a result, in general, the amplitude of oscillations at the same space and time is smaller when $d_{D,0}$ is smaller, see Figs. \ref{fig:numdenbase}. Especially when $d_{D,0}=0.001$, in this case, the reference Debye length is so small that ions and electrons rapidly combine and evolve like one fluid. The various coupling regimes are summarized in table~\ref{table-coupling}.

\begin{table}
\caption{Coupling regimes for various reference Debye length $d_{D,0}$.}
\begin{center}
\label{table_type}
\begin{tabular}{|c|c|c|}
\hline
~$d_{D,0}$~ &~ Extent of coupling ~& ~Behavior of ions and electrons~  \\
\hline
$\infty$ & No coupling & ~Two separate uncoupled hydro-fluids~  \\
$10$ & Loose coupling & ~Evolve almost as two separate hydro-fluids~  \\
$0.1$ & Weak coupling & ~The electrons oscillate around the ions due to the Lorentz force ~ \\
$0.01$ & Strong coupling & ~Similar as above with tighter combination and higher frequency~ \\
$0.001$ & ~Extreme strong coupling~ &  Ions and electrons evolve as one fluid    \\
$0$ & Full coupling & ~Equivalent to single-fluid MHD with $c\rightarrow\infty$ and $d_{D,0}c\rightarrow0$~ \\
\hline
\end{tabular}
\label{table-coupling}
\end{center}
\end{table}

From the above discussion, we note that the base states evolve like hydrodynamics when $d_{D,0}$ is either too large or too small. For a better understanding of the two-fluid effect, we investigate in detail the cases with finite $d_{D,0}$ ($0.1$ and $0.01$), i.e. the weak and strong coupling cases, in the ensuing sub-sections.

\subsection{\label{sec:3.1}Weak Coupling: \boldmath{$d_{D,0}=0.1$} Case}
Presently, we examine the linear dynamics due to the electron precursor waves, followed by an examination of the post-ion shock phase and then quantify the perturbation amplitude and growth rate history as a function of time.
\subsubsection{Electron Precursor Waves}
\begin{figure}[h]
\includegraphics[width=\linewidth]{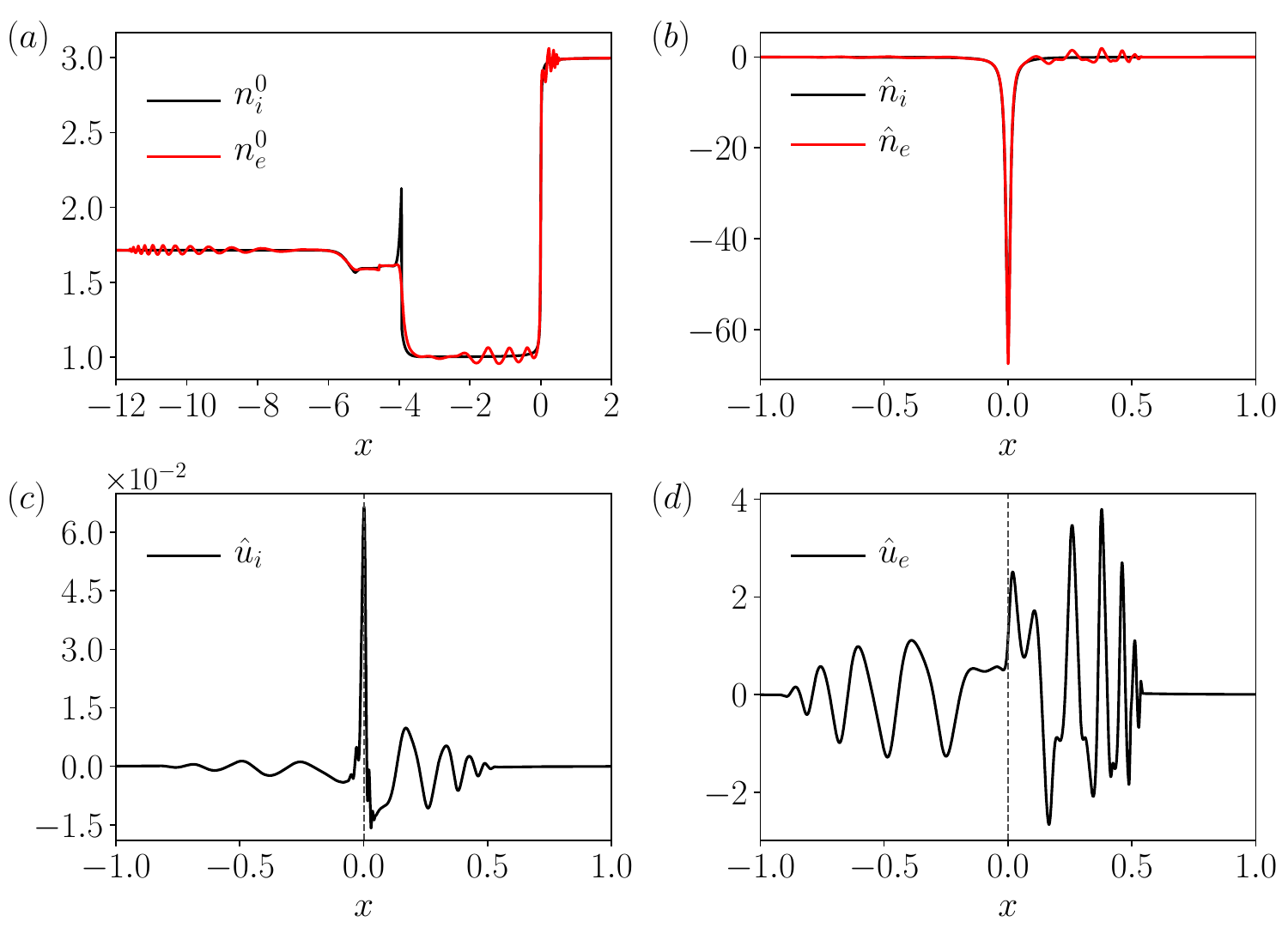}
\caption{\label{fig:uxpb_dd0_1_t0_628} Number density and velocity during the electron precursor waves interaction at $t=0.628$ for the weak coupling case with $d_{D,0}=0.1$; (a) number density profile of the base state, (b) perturbed number density state of ions and electrons, (c) perturbed ion velocity in $x$ direction, (d) perturbed electron velocity in $x$ direction. The dashed line denotes the location of the corresponding density interface. Only a limited portion of the entire domain $x \in [-1:1]$ is shown.}
\end{figure}

Figure \ref{fig:uxpb_dd0_1_t0_628} shows the number density and perturbed velocity for each species along the $x$ direction at $t=0.628$, where the $x-$ component of the perturbed velocity is derived as $\hat{u}_\alpha(x,t)=(\hat{\rho_\alpha u_\alpha}-u_\alpha^0\hat{\rho}_\alpha)/\rho_\alpha^0$. At this time, the electron waves have traversed the electron density interface due to the fast wave speed while the ion shock has only translated by a short distance (at $x \approx -4$). At the same time, the leftmost location of the electron waves {moves to $x\approx -12$ }(the {fast speed }of these waves {is }the reason for considering an infinite domain in $x$). The wave number of the electron waves increases across the electron density interface due to the high density there. Although not shown in the figure, the ion and electron density interfaces oscillate around initial position under the influence of the electron waves. For the perturbed states, we focus on a narrow region, $x \in [-1:1]$, near the density interface. Examining the perturbation equations, we note that any deviation of the base states from the equilibrium implies a forcing on the perturbed states. Therefore, the oscillating electron waves lead to an oscillating force, and this results in the oscillations  manifesting themselves in the perturbed states. Similar to the base state part, due to the high ion particle mass, the perturbed ion $\hat{n}_i$ and $\hat{u}_i$ are not as much influenced as electrons. For instance, the peak value of $\hat{u}_i$ is about $6.6\times10^{-2}$ while that of $\hat{u}_e$ is about $3.8$. 

\subsubsection{Post-Ion Shock Evolution}
\begin{figure}[h]
\includegraphics[width=\linewidth]{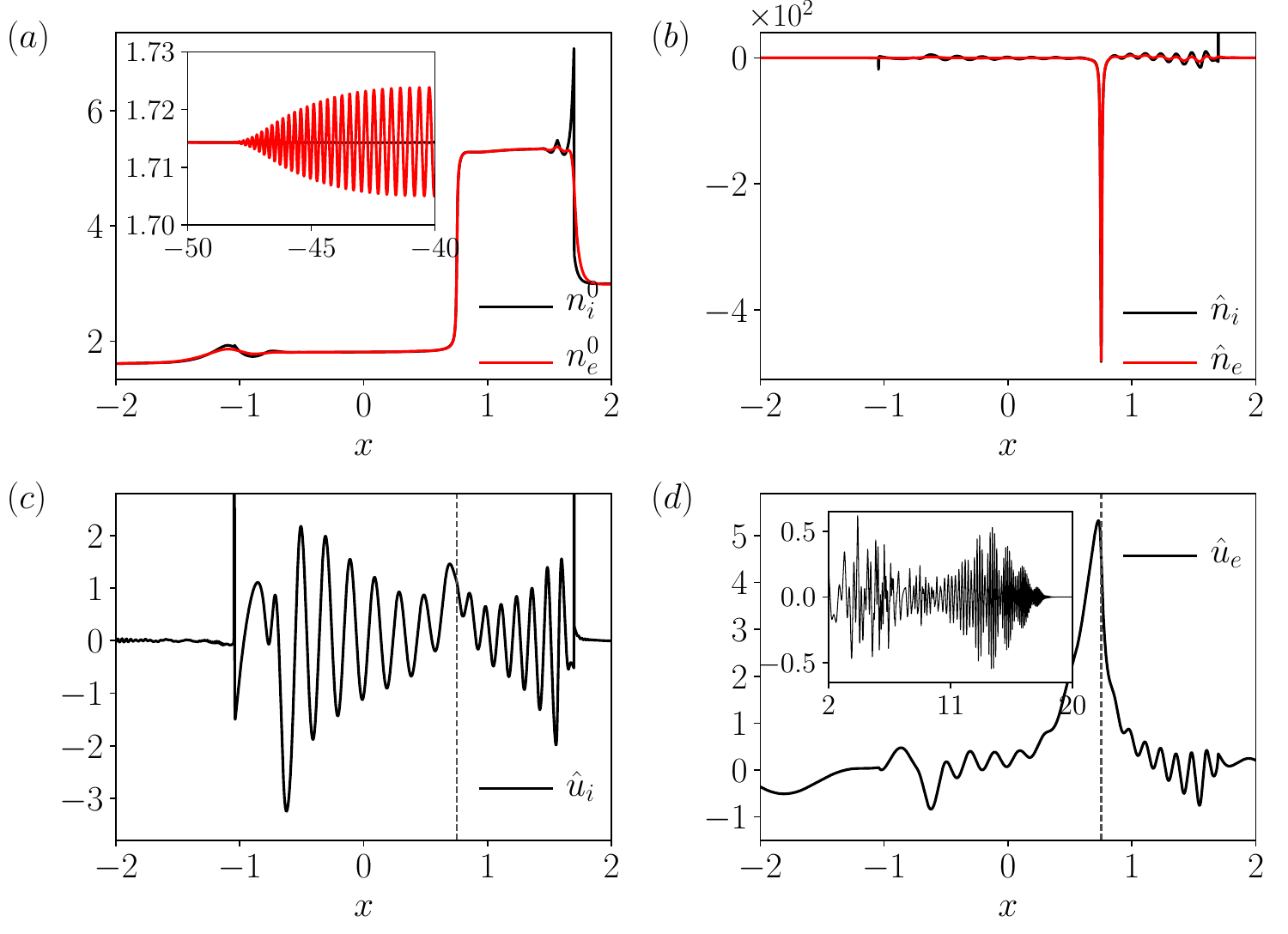}
\caption{\label{fig:uxpb_dd0_1_t4_241} Number density and velocity after the ion shock interaction with the interface at $t=4.241$ for the weak coupling case with $d_{D,0}=0.1$; (a) number density of base state (inset shows the location of the electron waves at $x\approx -48$), (b) number density of perturbed state, (c) perturbed ion velocity in $x$ direction, (d) perturbed electron velocity in $x$ direction. The dashed line denotes the location of the corresponding density interface.}
\end{figure}

At $t=4.241$, the ion shock is located at $x\approx1.7$ and has interacted with the ion interface, as plotted in Fig. \ref{fig:uxpb_dd0_1_t4_241}. The ion shock-interface interaction breaks the oscillation motion of the ion density interface around $x=0$ by providing an impulse,  so that the ion interface moves to the location $x\approx0.75$ at the time instant shown. The induced electron Lorentz force drags the electron density interface to evolve together with ion interface. Meanwhile, the leftmost position of the electron waves is about $-48$ (inset in Fig.~\ref{fig:uxpb_dd0_1_t4_241}(a)). Unlike the weak electron waves, the ion shock can significantly change the base states of the two species, so that substantial forces are exerted on both ion and electron perturbations. Therefore, the perturbed states in the ions deviate sufficiently enough that both $\hat{n}_i$ and $\hat{u}_i$ are comparable to the electron perturbations. It is the perturbed velocity $\hat{u}_i$ at the interface that is related to the growth rate of the perturbation amplitude. Ideally, if the ion shock is a discontinuity, it will result in delta-function like spike in the perturbation of the number density and momentum at the ion shock location. However, numerically due to the smearing of the shock front,  there is an extremely large spike in $\hat{n}_i$ and $\hat{u}_i$ at the location of ion shock. Note that the peaks {of these} spikes are actually larger than shown in Fig.\ref{fig:uxpb_dd0_1_t4_241}(c) and these peaks have been cut off as these isolated peaks are not important to our discussion.

\subsubsection{Perturbation Amplitude and Growth Rate History}
\begin{figure}[h]
\includegraphics[width=\linewidth]{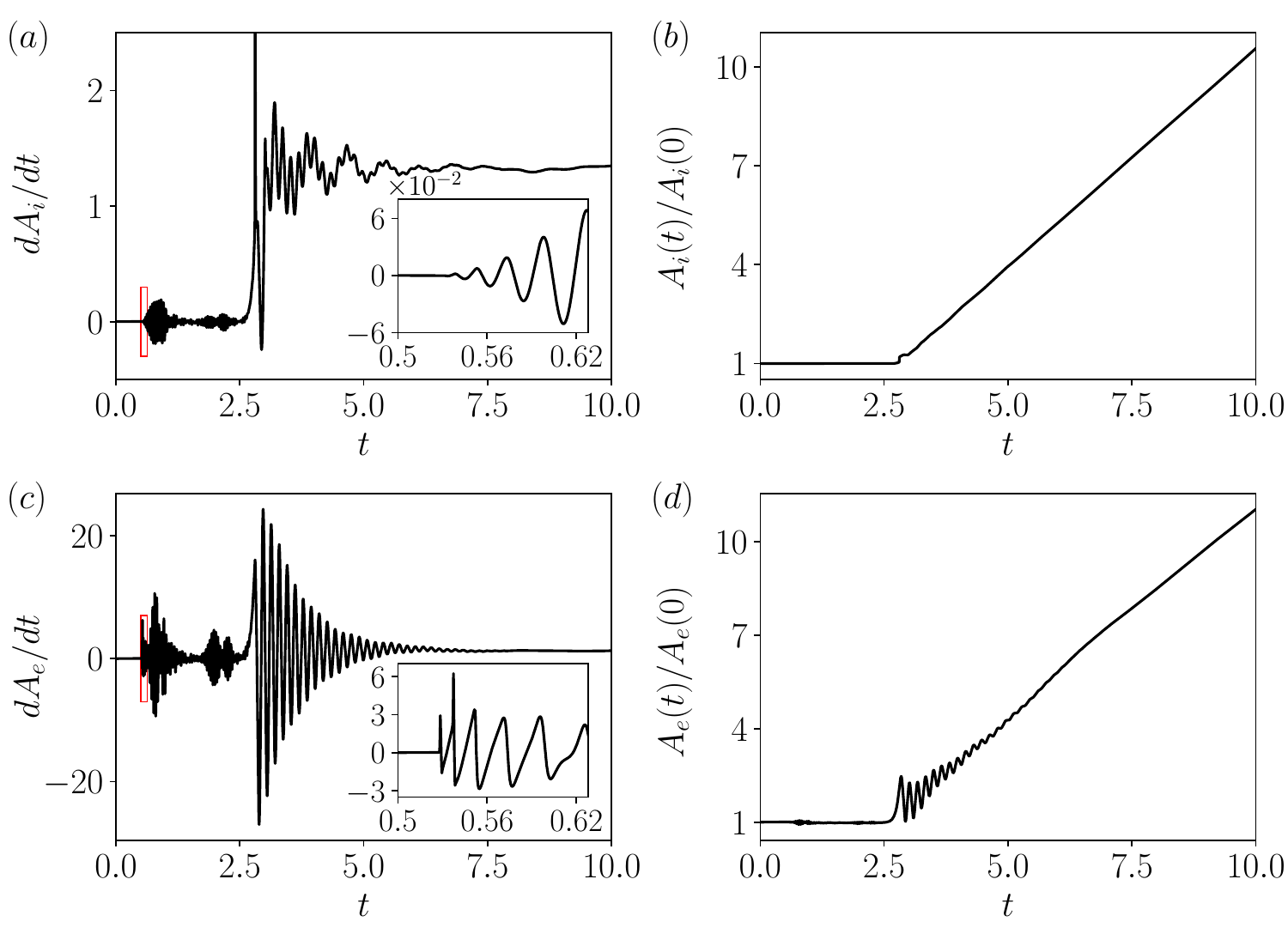}
\caption{\label{fig:gr_dd0_1_binf} Evolution of the growth rate and reference amplitude of the ion and electron density interfaces for the weak coupling case with $d_{D,0}=0.1$; (a) growth rate of ion interface, (b) amplitude of ion interface, (c) growth rate of electron interface, (d) amplitude of electron interface. }
\end{figure}

We now discuss the time history of the perturbation amplitude and its growth rate for both the ion and electron species. The location of each interface at all times is determined by tracking the tracer variable of each species. In our simulations, for each species, the value of tracer $\phi_\alpha$ is set to be $-1$ (respectively $+1$) on the left (respectively, right) of the corresponding density interface. Thus, the growth rate of each interface (denoted as $dA_\alpha/dt$ in the figures) is computed as the perturbed velocity $\hat{u}_\alpha$ where $\phi_\alpha(x,t)=0$. The evolution of perturbation amplitude for each species is then calculated by integrating the growth rate over time. Figure \ref{fig:gr_dd0_1_binf} shows the evolution of growth rate and amplitude of the perturbations at each density interface. The precursor electron waves first interact with the electron interface at about $t=0.53$, while the ion shock-interface interaction is delayed and occurs around $t=2.7$. As discussed above, the electron waves induce an oscillating force on the perturbation states, giving rise to the oscillating perturbed velocities of the interfaces. Thus, the growth rate of each interface oscillates during the interaction between electron waves and interface, i.e., oscillatory growth rate in both the ions and electrons due to the electron precursor waves occurs during the interval $t\in[0.53,2.7]$. During this interval, although the amplitude of the oscillating electron growth rate is considerable, the period of each cycle is so short that the integral of positive growth rate over the half period is not large and quickly decreases in the other half period of negative growth rate. As shown in Fig. \ref{fig:gr_dd0_1_binf}(d) the electron perturbation amplitude $A_e$ oscillates with very small amplitude before $t=2.7$,  and we note from Fig. \ref{fig:gr_dd0_1_binf}(b) that the  ion perturbation amplitude $A_i$ appears to be virtually zero although in actuality it is oscillating with a much smaller amplitude. 

The dynamics  become quite different after the ion shock impacts the ion interface. The substantially altered ion base state exerts a considerable influence on the ion perturbed state. As a result, the growth rate of ion interface increases very rapidly in a very short period during the ion shock-interface interaction, which is seen as a sharp  spike at $t\approx 2.7$.  After the interaction, the ion interface moves right with a positive base velocity $u_{i,I}^0$ while the amplitude grows with a positive perturbed velocity $\hat{u}_{i,I}$. Again, the Lorentz force exerted on the perturbations tends to oscillate the growth rate, however the force is not enough to change the sign of $dA_i/dt$. At late time, $dA_i/dt$ gradually oscillates around $1.32$ -- this is somewhat similar to the growth rate noted in linearized compressible hydrodynamics as originally done by Richtmyer {\citep{richtmyer1960}}. The amplitude, $A_i$, appears to grow linearly in time. In contrast to the ions, the force on electron perturbations are significantly affected and $dA_e/dt$ can even drop below zero during a short duration after $t=2.7$. Therefore, we can see an obvious oscillation in the $A_e$ during the time interval from $t\in[2.7,5.0]$. As time increases, the decreasing difference between base states diminishes the force acting on the perturbations. As a consequence, the force eventually becomes too weak to significantly influence the electron growth rate at late time. Meanwhile, $dA_e/dt$ gradually oscillates around $1.28$.

It is interesting to observe that the different frequencies in the growth rate plots. We investigate the frequencies by dividing the time zone into two intervals: $T_A$ ($0.53 < t < 2.7$) and $T_B$ ($2.8 < t <10$).  During the time interval $T_A$, the growth rate is caused by the interaction between the precursor electron waves and interface, thus the frequency $\Omega_e$ is related to the frequency of electron waves, i.e.,  $\Omega_e \sim \frac{1}{d_{D,0}}\sqrt{\frac{{n_e}}{m_e}}$. During the time interval $T_B$, two frequencies are observed: the large one which is also clearly seen in $dA_e/dt$ is linked to the frequency of ion waves, so that $\Omega_i \sim \frac{1}{d_{D,0}}\sqrt{\frac{{n_i}}{m_i}}$. The smaller frequency,  one that is very noticeable in $dA_i/dt$, corresponds to the reverberations, denoted as $\Omega_r$. These reverberations are also present in the linearized solution to compressible hydrodynamics equations, and are related to the sound waves reverberations in the $y-$ direction {\citep{yang1994}}. This reverberation frequency {$\Omega_r$ is not related to the parameter $d_{D,0}$ when $d_{D,0}$ is sufficient small}. Generally, these frequencies satisfy $\Omega_e>\Omega_i>\Omega_r$. It is noted that all three frequency waves exist and are superposed in both of the growth rate plots. However, the oscillations at the $\Omega_e$ frequency are too weak to be distinct in the plots after $t>2.7$. 

\subsection{\label{sec:3.2}Strong Coupling: \boldmath{$d_{D,0}=0.01$} Case}
\begin{figure}
\includegraphics[width=\linewidth]{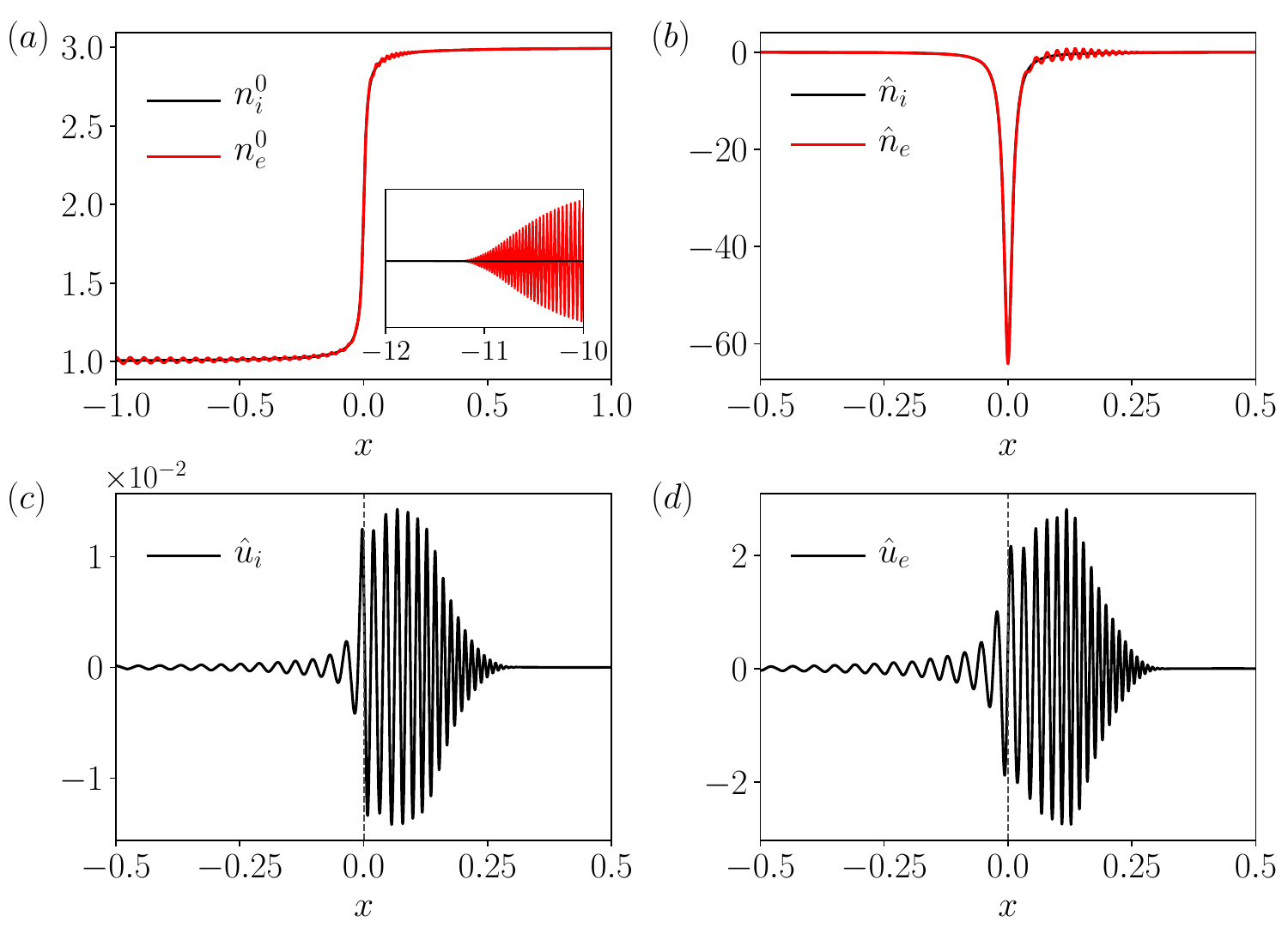}
\caption{\label{fig:uxpb_dd0_01_t0_628} Number density and velocity during the electron precursor interactions at $t=0.628$ for the strong coupling case with $d_{D,0}=0.01$; (a) number density of base state, (b) number density of perturbed state, (c) perturbed ion velocity in $x$ direction, (d) perturbed electron velocity in $x$ direction. The dashed line denotes the location of the corresponding density interface. }
\end{figure}

\begin{figure}
\includegraphics[width=\linewidth]{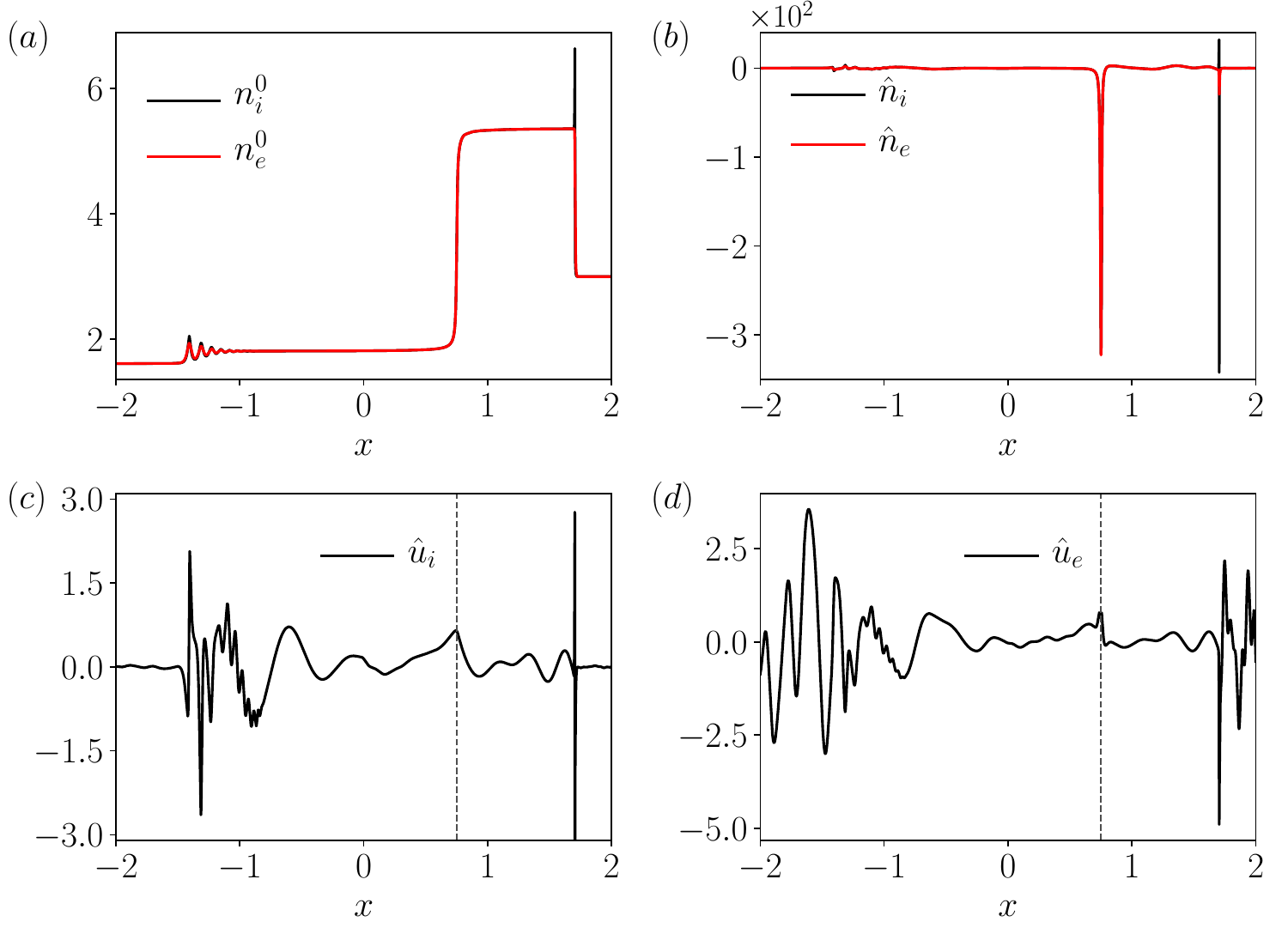}
\caption{\label{fig:uxpb_dd0_01_t4_241} Number density and velocity after the ion-shock interaction with the interface at $t=4.241$ for the strong coupling case with $d_{D,0}=0.01$; (a) number density of base state, (b) number density of perturbed state, (c) perturbed ion velocity in $x$ direction, (d) perturbed electron velocity in $x$ direction. The dashed line denotes the location of the corresponding density interface. }
\end{figure}

\begin{figure}
\includegraphics[width=\linewidth]{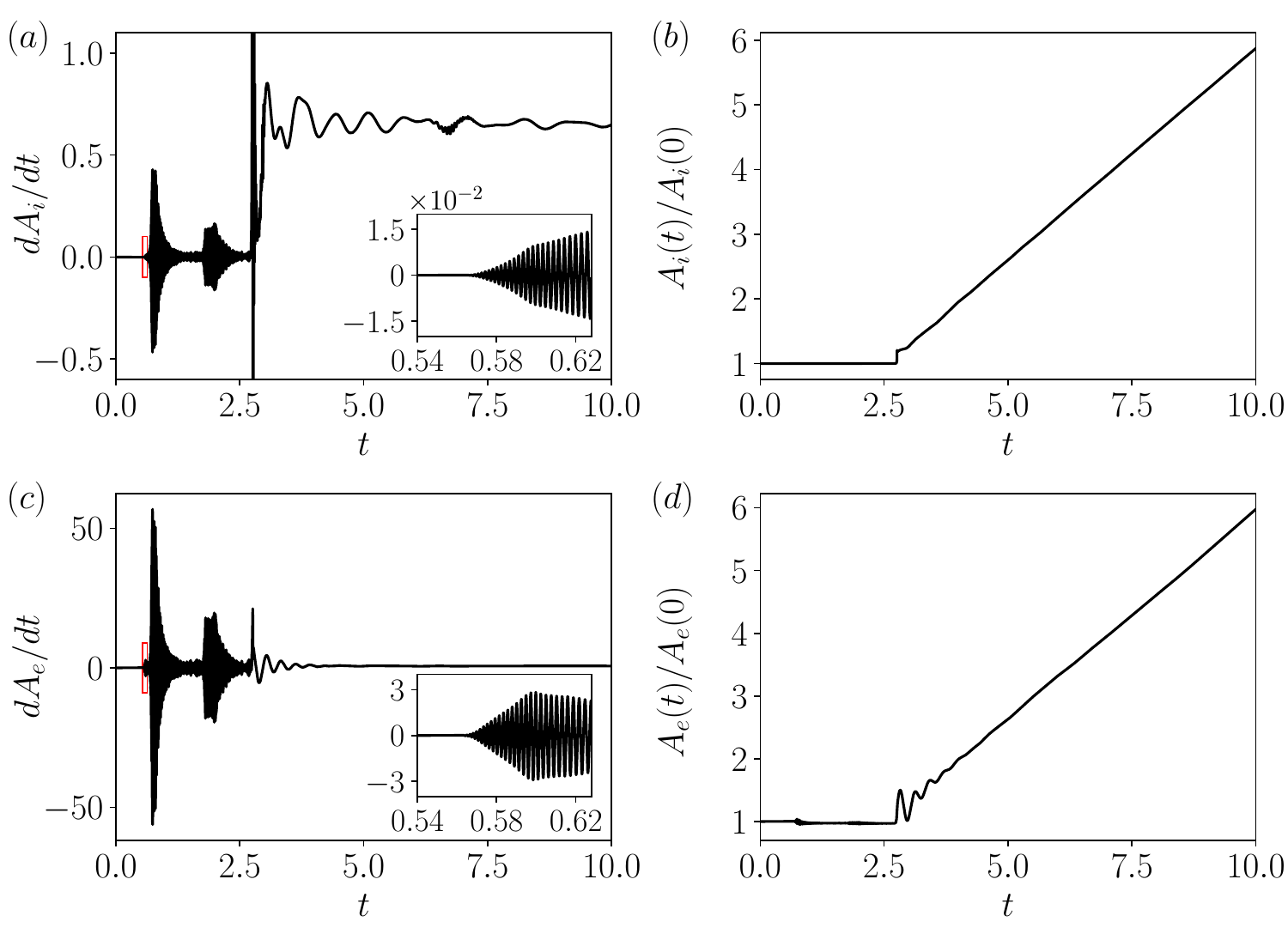}
\caption{\label{fig:gr_dd0_01_binf} Evolution of the growth rate and perturbation amplitude of the ion and electron density interfaces for the strong coupling case with $d_{D,0}=0.01$; (a) growth rate of ion interface, (b) amplitude of ion interface, (c) growth rate of electron interface, (d) amplitude of electron interface. }
\end{figure}
In this case, the smaller $d_{D,0}$ implies a stronger coupling between the two charged species compared to the previous one. Thus, the motions of ions and electrons are expected to be more similar to each other. 
As before, we examine the linear dynamics due to the electron precursor waves, followed by an examination of the post-ion shock phase and then quantify the perturbation amplitude and growth rate history as a function of time.
\subsubsection{Electron Precursor Waves}
Fig. \ref{fig:uxpb_dd0_01_t0_628} shows the number density and perturbed velocity for each species along the $x$ direction at $t=0.628$. We can see the base number densities match well with each other. The electron number density shows oscillations with a higher wavenumber but with a smaller amplitude compared with $d_{D,0}=0.1$ case. At this time, the leftmost location that electron waves have reached at $x\approx -11.3$, closer than that of the $d_{D,0}=0.1$ case, and is a consequence of the coupling effect that tends to accelerate ions while decelerating the electrons. Thus, the smaller $d_{D,0}$ is correlated with the the slower electron wave speed. Similar to the $d_{D,0}=0.1$ case, after the interaction between the precursor electron waves and the interface, the electron base state induces an oscillating force on the electron perturbations, results in the oscillating perturbations (see, for instance, the $\hat{n}_e$ and $\hat{u}_e$ in the Fig.\ref{fig:uxpb_dd0_01_t0_628}). The deviated electron base state induces the Lorentz force acting on the ions that only slightly changes the ion base state due to their large particle mass. However, the change is smaller compared to the electron part and the magnitude of $u_i$ is about two orders smaller than the magnitude of $\hat{u}_e$. However, the profiles of perturbed velocities of ions and electrons are virtually similar (insofar as the location and wavenumber is concerned) for this smaller  $d_{D,0}=0.01$ case than for the $d_{D,0}=0.1$ case. 

\subsubsection{Post-Ion Shock Evolution}
At $t=4.241$  the ion shock has already impacted the ion interface (see Fig. \ref{fig:uxpb_dd0_01_t4_241}). Due to the stronger coupling effect, the base number densities of ions and electrons profiles in $x$ are virtually  identical  compared with the $d_{D,0}=0.1$ case. The ion shock strongly changes not only the ion base state but also the electron one. As a result, the perturbation states of both species are strongly influenced by the ion shock. We note that the order of magnitude of perturbed ion $x-$ component of velocity $\hat{u}_i$ is same with that of $\hat{u}_e$. Furthermore, the structures of $\hat{u}_i$ and $\hat{u}_e$ are more alike than that of the large $d_{D,0}$ case.

\subsubsection{Perturbation Amplitude and Growth Rate History}
The time history of the growth rate and amplitude of perturbation for both species are plotted in Fig. \ref{fig:gr_dd0_01_binf}. After the first interaction between precursor electron waves and the interface at $t\approx 0.56$, the electron perturbations start to grow. Due to the oscillating nature of the electron waves, the growth rate of electron density interface $dA_e/dt$ oscillates around zero with frequency $\Omega_e \sim \frac{1}{d_{D,0}}\sqrt{\frac{n_e}{m_e}}$. Due to the drag effect of the Lorentz force, the ion growth rate $dA_i/dt$ also oscillates around zero with same frequency but the amplitude is about two orders of magnitude smaller than the electrons.  At early time, because the perturbed Lorentz force is inversely proportional to $d_{D,0}$, the perturbations are subject to a larger force in the smaller $d_{D,0}$ case, and hence this results in the larger amplitude of the growth rate. We note that the peak amplitude of $dA_e/dt$ is about $57$ in the case with $d_{D,0}=0.01$ while is about $11$ in the $d_{D,0}=0.1$ case. As time progresses, the Lorentz force induced due to the difference in the base state, tends to reduce these difference, and reduce the forcing source term on the perturbed quantities. As a result, the amplitude of the oscillating growth rate gradually decreases, as shown in the figure. Though the magnitude of the oscillating growth rate is considerable, the short period of each cycle leads to the amplitude of the perturbations almost unchanged over time, similar as in the $d_{D,0}=0.1$ case. The ion growth rate $dA_i/dt$ develops significantly after the  ion shock-interaction at $t\approx 2.6$. {On} one hand, the smaller $d_{D,0}$ implies a larger force acting on the perturbations. On the other hand, the smaller $d_{D,0}$ also implies a shorter period for the base state difference, thus the shorter period for the force affects the perturbations. Hence, a competitive mechanism develops. In this case, the latter one dominates so that $dA_i/dt$ increases to a value smaller than that of case with $d_{D,0}=0.1$. At late time, the ion growth rate oscillates around $0.65$ while the electron growth rate oscillates around $0.66$. The normalized amplitude of ion perturbations is about $5.87$ while that of electron perturbations is about $5.98$ at $t=10$. For comparison, for the $d_{D,0}=0.1$ case, the reference amplitudes of ion and electron interfaces at $t=10$ is about $10.56$ and $11.04$, respectively. Since the frequencies $\Omega_e$ and $\Omega_i$ are inversely proportional to $d_{D,0}$, these frequencies are much larger than ones for the $d_{D,0}=0.1$ case, as confirmed in Fig. \ref{fig:gr_dd0_01_binf}. However, the frequency $\Omega_r$ is comparable between the two cases. It is noted that the oscillations with frequency $\Omega_\alpha$ are present in the growth rate for  $t>2.6$, although these are too weak to be visible in the plots. 

\section{\label{sec:4} Effect of the Initial Magnetic Field}
\begin{figure}
\includegraphics[width=\linewidth]{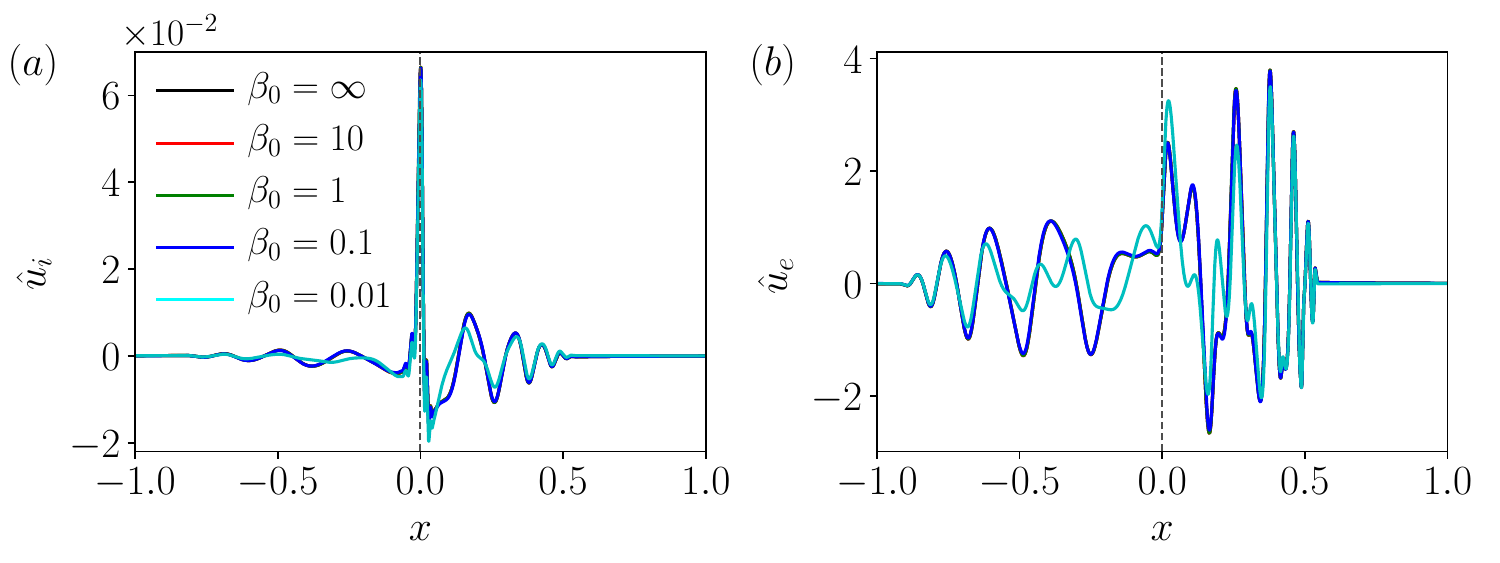}
\caption{\label{fig:uxpb_dd0_1_mag} Perturbed velocity (x-component) of ions and electrons for various $\beta_0$ at $t=0.628$; the reference Debye length $d_{D,0}=0.1$. (a) ions: $\hat{u}_i$, (b) electrons: $\hat{u}_e$. }
\end{figure}

\begin{figure}
\includegraphics[width=\linewidth]{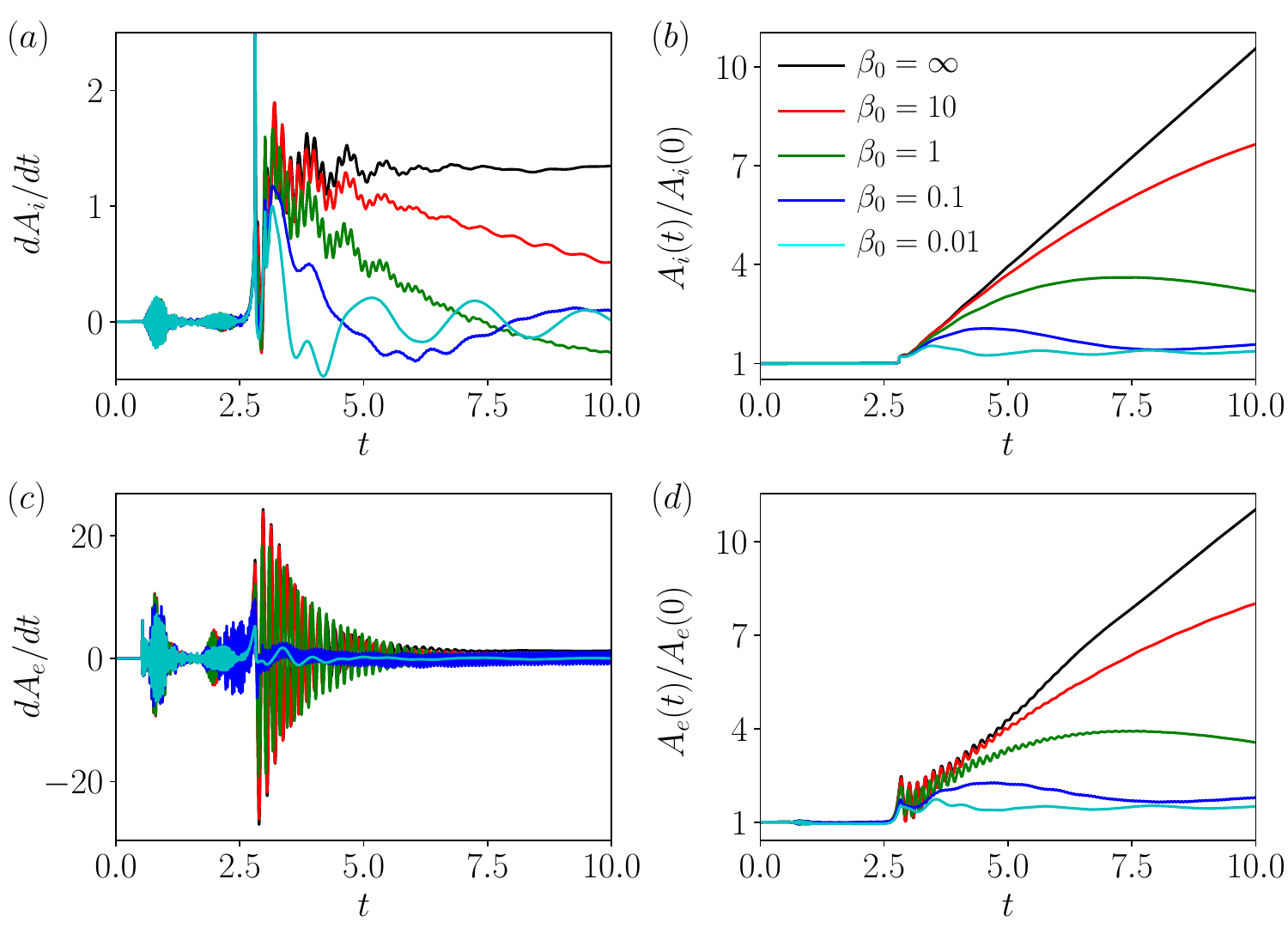}
\caption{\label{fig:gr_dd0_1_mag} Evolution of the growth rate and reference amplitude of the density interfaces for various $\beta_0$.  The reference Debye length $d_{D,0}=0.1$. (a) growth rate of ion interface, (b) amplitude of ion interface, (c) growth rate of electron interface, (d) amplitude of electron interface. }
\end{figure}

\begin{figure}
\includegraphics[width=\linewidth]{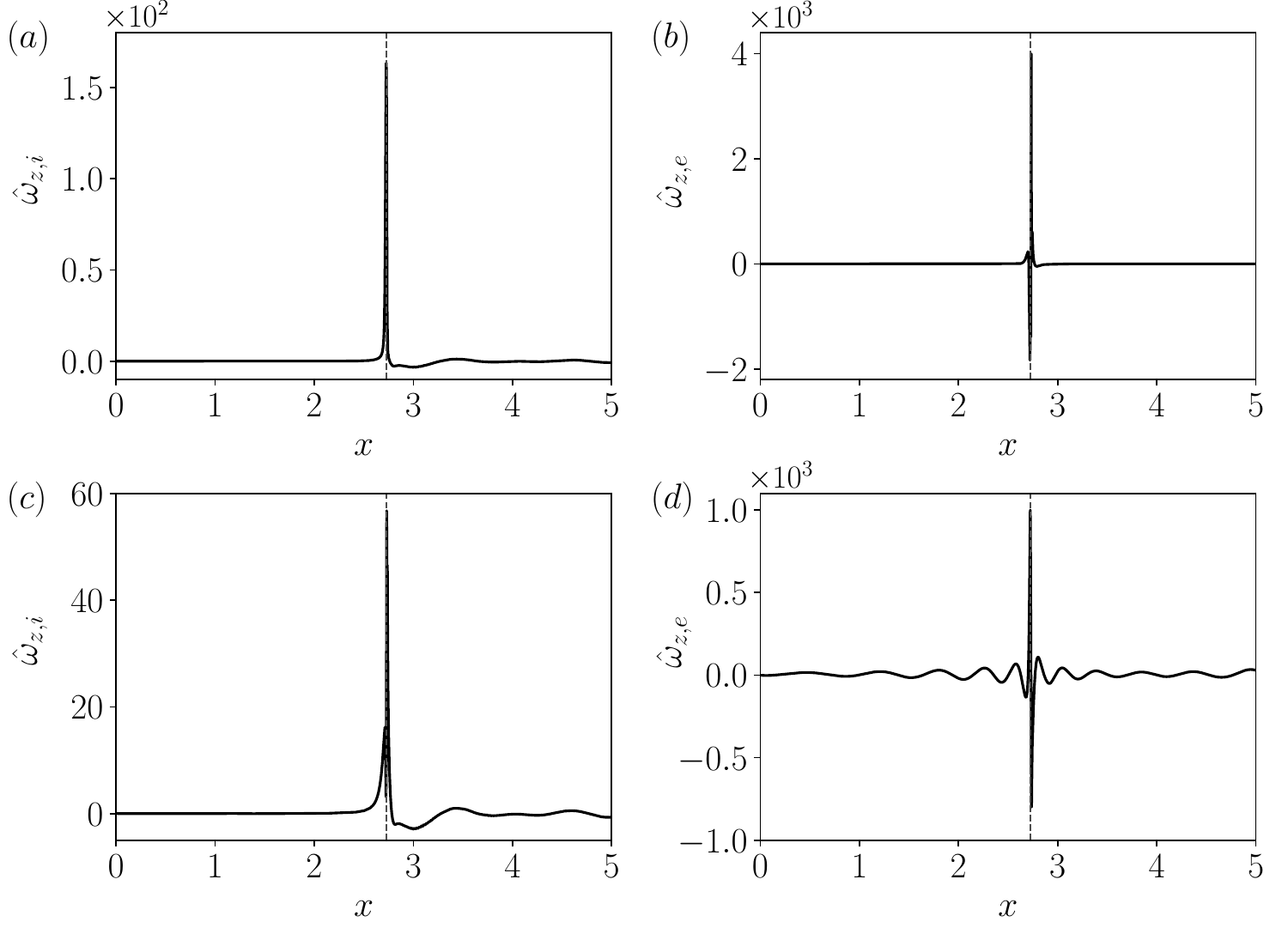}
\caption{\label{fig:dd0_1_vor_t8_168} Perturbed vorticity of ions and electrons in $z$ direction at $t=8.168$ for the cases with $\beta_0=$ $\infty$(a, b) and $10$ (c, d). The reference Debye length $d_{D,0}=0.1$. The dashed line denotes the location of the corresponding density interface.}
\end{figure}

\begin{figure}
\includegraphics[width=\linewidth]{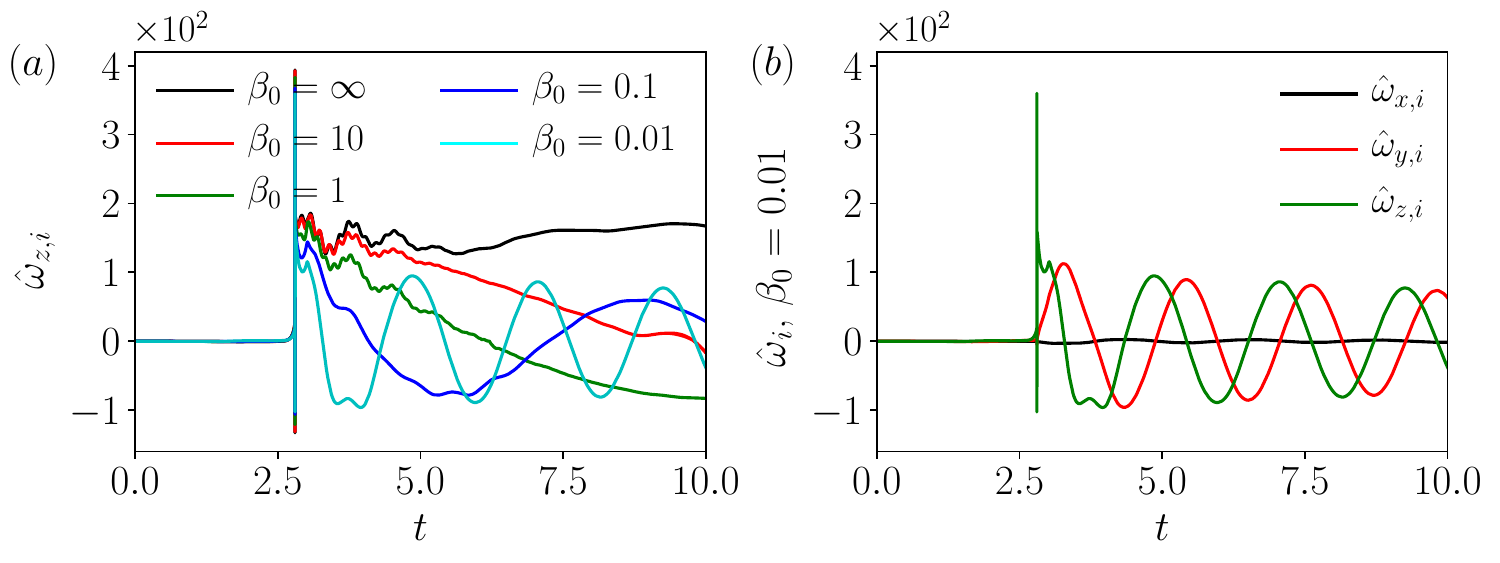}
\caption{\label{fig:gr_dd0_1_vor} (a) Evolution of perturbed ion vorticity in $z$ direction on the ion interface for the cases with various $\beta_0$, (b) evolution of each component of perturbed ion vorticity on the ion interface for the case with $\beta_0=0.01$ . The reference Debye length $d_{D,0}=0.1$.}
\end{figure}
\subsection{\label{sec:4.1} Weak Coupling: \boldmath{$d_{D,0}=0.1$} Case}
Presently, we turn our attention to the effect of the initial imposed magnetic field on the growth of the perturbations for the weakly coupled, i.e., $d_{D,0}=0.1$, case. 
%In the context of incompressible ideal MHD, Wheatley \etal {\citep{wheatley2005}} showed that the initial growth rate matches well the impulse model of Richtmyer and eventually the growth rate decreases to zero with a corresponding saturation of the perturbation amplitude. Hence the application of the magnetic field is associated with a time delay before the growth rate shrinks to zero. 
In the two-fluid plasma model, there is a time delay before the magnetic field influence is felt on the perturbed quantities. 
This time delay before the magnetic field has an influence is proportional to $\beta_0$, i.e., the stronger (small $\beta_0$)  the field the shorter the delay. 
This delayed influence of the magnetic field is illustrated by examining the flow field at early time ($t=0.628$) {in Fig.} \ref{fig:uxpb_dd0_1_mag} where the the perturbed velocities are plotted for different $\beta_0$. 
The influence of initial $x-$ direction magnetic field on the flow continues to grow with time. 
At this early time, the  effect of the magnetic field with $\beta_0\leq0.1$ is not sufficient enough to affect perturbed velocities, while the magnetic field with $\beta_0=0.01$ apparently changes $\hat{u}_\alpha$, especially for the $\hat{u}_e$ due to the light particle mass. 

%Therefore, even the strong magnetic field (small $\beta_0$) has little influence on the flow at the very early time, while the weak magnetic field (large $\beta_0$) may have significant effect on the flow at the very late time.

The time history of growth rate and normalized perturbation for ions and electrons are plotted in Fig. \ref{fig:gr_dd0_1_mag}. During the electron precursor wave interaction, the growth rate is virtually the same for all $\beta_0$ except for the strongest field case ($\beta_0=0.01$) which shows a small decrease in the growth rate. 
Just after the ion shock interaction with the interface, the growth rate and perturbation amplitude matches well with each other for various $\beta_0$.  As time progresses we see differences in the growth rates for different values of  $\beta_0$. For ions, the peak value of growth rate $dA_i/dt$ induced by the ion shock decreases as the strength of the magnetic field increases (see Fig. \ref{fig:gr_dd0_1_mag}(a)). For $\beta_0< 1$, the ion growth rate is significantly suppressed by the magnetic field and, in fact, the growth rate dips below zero.  The stronger the initial magnetic field, the faster the ion growth rate decreases below zero. After that,  $dA_i/dt$ oscillates around zero with a frequency that is proportional to the ion cyclotron frequency, i.e., $\Omega_B \sim \frac{1}{m_i d_{D,0}\sqrt{\beta_0}}$, and this frequency is larger for stronger magnetic fields.  For $\beta_0=0.01$, there are about three cycles (or periods) that are captured within the simulation duration, while for $\beta_0=0.1$ we note only about one such cycle. For the case with $\beta_0=1$ or $10$, the frequency of the cycle is smaller than the duration of the simulation and only a part of this oscillation cycle is observed. The perturbation amplitude is computed by integrating the growth rate over time (see  Fig. \ref{fig:gr_dd0_1_mag}(b)). The ion perturbation amplitude is also suppressed by the magnetic field after a short duration during which the amplitude grows. Here it is relevant to point out that, in the context of incompressible ideal MHD, Wheatley \etal {\citep{wheatley2005}} showed that the initial growth rate matches well the impulse model of Richtmyer and eventually the growth rate decreases to zero with a corresponding saturation of the perturbation amplitude. We note a somewhat similar trend here, i.e., the interface amplitude grows for a short duration after which the magnetic field influences the dynamics, reduces the growth rate, and the perturbation amplitude is smaller than it would be without the initial magnetic field.   Hence the application of the magnetic field is associated with a time delay before the growth rate shrinks to zero. The stronger the initial magnetic field, the larger is the extent the suppression. Since the ion growth rate eventually oscillates around a zero mean value, the ion perturbation amplitude would oscillate around a finite value at the end. 
The electron growth rate and perturbation amplitude is also suppressed by the magnetic field as seen in Fig. \ref{fig:gr_dd0_1_mag}(c) and (d), respectively. The mechanisms affecting the electron dynamics are similar to that influencing the ions, but the electrons respond much faster to the magnetic field with a higher oscillation frequency owing to their lighter mass. For the electrons, several oscillation cycles about zero mean are noted even for higher values of $\beta_0$. The suppression mechanism is further discussed next by examining the vorticity evolution on the interface. 

In RM instability in hydrodynamics, as the shock interacts with the interface, baroclinic generation of vorticity occurs which drives the growth of the perturbations.  Similarly, we expect that the vorticity on the interface is the driving force which results in the growth of the perturbations. The relevant quantity in linear analysis is the $z$ component of the perturbed vorticity defined as $\hat{\omega}_{z,\alpha}=Re(d\hat{v}_\alpha/dx-ik\hat{u}_\alpha)$. Figure \ref{fig:dd0_1_vor_t8_168} compares the perturbed vorticity of ions and electrons with or without initial magnetic field at $t=8.168$. We see that the vorticity on the interface of each species reduces in the presence of magnetic field. {For instance, the vorticity on the ion interface is $\hat{\omega}_{z,i}\approx 160.23$ when $\beta_0=\infty$ while  $\hat{\omega}_{z,i}\approx 26.08$ when $\beta_0$ is $10$. We further note that the peak vorticity may not be coincide with the location of the interface}. It shows that the vorticity on the interface is transported away when the initial magnetic field is applied. As a consequence, the growth rate of the perturbations are suppressed by the field. 

{Figure \ref{fig:gr_dd0_1_vor}(a) shows the evolution of $\hat{\omega}_{z,i}$ on the ion interface (where $\phi_i(x,t)=0$) for various $\beta_0$. After the ion shock-interface interaction at $t\approx 2.7$, vorticity $\hat{\omega}_{z,i}$ is deposited on the interface. After that, $\hat{\omega}_{z,i}$ is positive when $\beta_0=\infty$ or oscillates around $0$ with frequency proportional to $\Omega_B$ when $\beta_0$ is finite. The time history of  $\hat{\omega}_{z,i}$ is strongly correlated with the ion growth rate $dA_i/dt$ of the ion interface (see Fig. \ref{fig:gr_dd0_1_mag}). This is consistent with vortex dynamical interpretation of RM instability \citep{zabusky1999}. The oscillating $\hat{\omega}_{z,i}$ leads to the oscillating growth rate, and results in the overall suppression of RM instability. When the initial magnetic field is applied, the perturbed Lorentz force $\hat{\mathcal{L}}_i$ becomes considerable enough to transport the vorticity away from the interface. 
The effect of $\nabla \hat{\mathcal{L}}_i$ ($\hat{\mathcal{L}}_i$) is to decrease the magnitude of perturbed ion vorticity $\hat{\bm{\omega}}_{i}$ (velocity $\hat{\bm{u}}_i$). However, there is an out-of-phase correlation between between $\nabla \hat{\mathcal{L}}_i$ ($\hat{\mathcal{L}}_i$) and $\hat{\bm{\omega}}_{i}$ ($\hat{\bm{u}}_i$). The Lorentz force changes sign and  leads to the overshoot of $\hat{\bm{\omega}}_{i}$ ($\hat{\bm{u}}_i$). Eventually the cycle repeats and as a result, oscillation occurs in vorticity, and correspondingly the growth rate oscillates. As shown in Fig. \ref{fig:gr_dd0_1_vor}(b), each component of $\hat{\bm{\omega}}_{i}$ oscillates around zero, which is also observed in the nonlinear two-fluid plasma simulation result \citep{bond2020}. The same process occurs in the electrons though not elaborated here.}
%The effect of $\nabla \hat{\mathcal{L}}_i$ ($\hat{\mathcal{L}}_i$) is to decrease the magnitude of perturbed ion vorticity $\hat{\bm{\omega}}_{i}$ (velocity $\hat{\bm{u}}_i$). However,  the out of phase between $\nabla \hat{\mathcal{L}}_i$ ($\hat{\mathcal{L}}_i$) and $\hat{\bm{\omega}}_{i}$ ($\hat{\bm{u}}_i$), leads to the overshoot of $\hat{\bm{\omega}}_{i}$ ($\hat{\bm{u}}_i$) and forms one circle when the overshoot turns back. As a result, oscillation occurs in vorticity (growth rate). As shown in Fig. \ref{fig:gr_dd0_1_vor}(b), each component of $\hat{\bm{\omega}}_{i}$ oscillates around zero, which is also observed in the nonlinear two-fluid plasma simulation result \citep{bond2020}. The same process occurs in the electrons though not illustrated here.  }

\subsection{\label{sec:4.1} Strong Coupling: \boldmath{$d_{D,0}=0.01$} Case}
\begin{figure}[h]
\includegraphics[width=\linewidth]{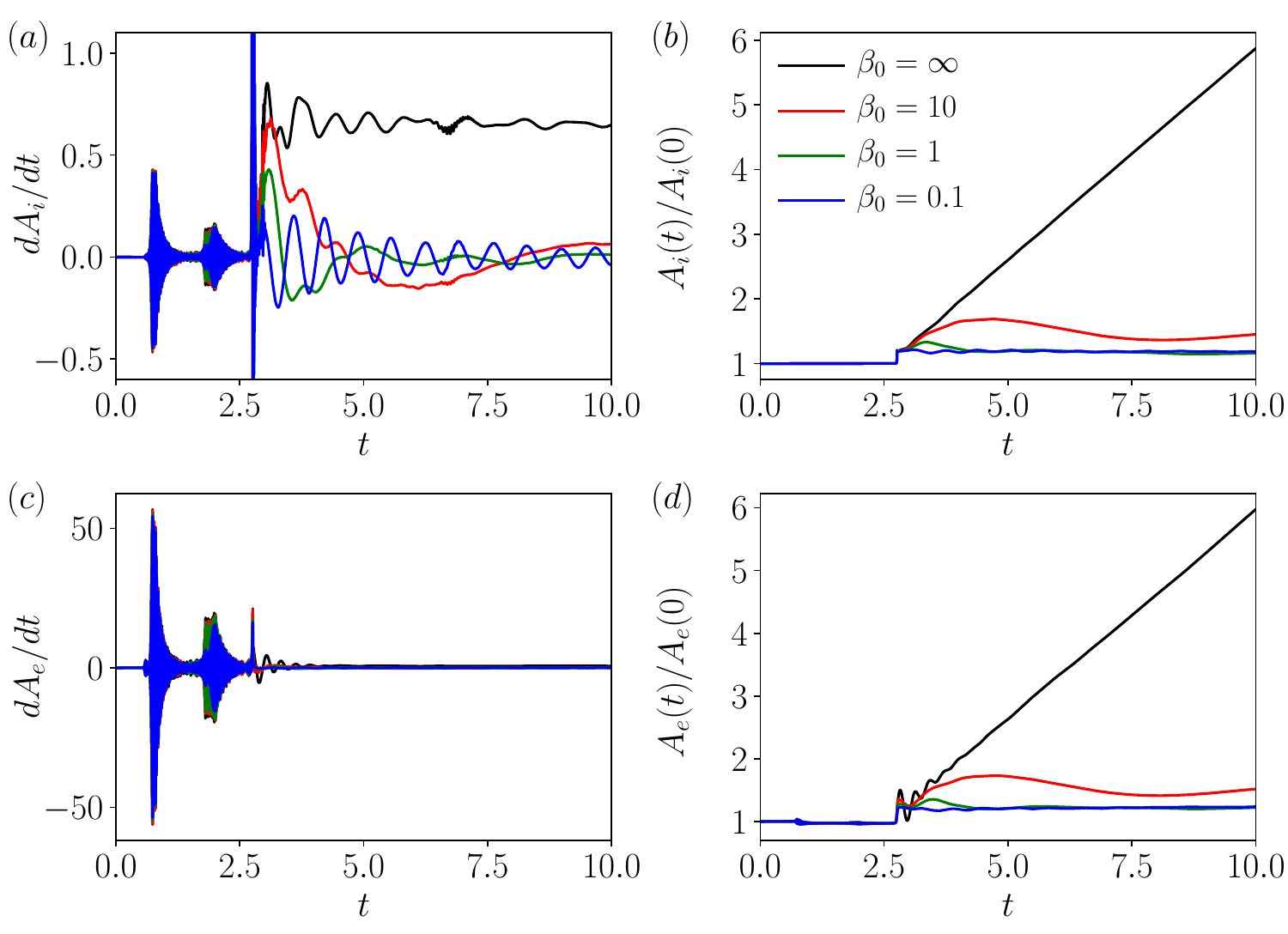}
\caption{\label{fig:gr_dd0_01_mag} Evolution of the growth rate and reference amplitude of the density interfaces for the cases with various $\beta_0$; the reference Debye length $d_{D,0}=0.01$. (a) growth rate of ion interface, (b) amplitude of ion interface, (c) growth rate of electron interface, (d) amplitude of electron interface. }
\end{figure}
For this strong coupling case, the time history of growth rate and normalized perturbation for ions and electrons are plotted in Fig. \ref{fig:gr_dd0_01_mag}. At early time before the ion shock interaction with the interface, the magnetic field has little influence on the perturbations, and the growth rate and perturbation amplitude match well with each other for various $\beta_0$. This is similar to the weak coupling case. As time progresses, the growth rate is significantly suppressed by the magnetic field with various $\beta_0$. For ions, the peak value of growth rate $dA_i/dt$ induced by the ion shock decreases as the strength of the magnetic field increases (see Fig. \ref{fig:gr_dd0_01_mag}(a)). In addition, the ion growth rate dips below zero and then $dA_i/dt$ oscillates around zero with a frequency that is proportional to the ion cyclotron frequency, i.e., $\Omega_B \sim \frac{1}{m_i d_{D,0}\sqrt{\beta_0}}$. The stronger the magnetic field, the larger is the frequency of the oscillation (or the shorter period for ion growth rate to dip to zero). For $\beta_0=0.1$, we observe ten cycles that are captured within the simulation duration ($1$ cycle for the $d_{D,0}=0.1$ case), while for $\beta_0=1$ we note only about three such oscillation cycles, and for the case with $\beta_0=10$  there is about one oscillation cycle. Since $\Omega_B$ is inverse proportional to $d_{D,0}$, the frequency of the oscillating growth rate for $d_{D,0}=0.01$ case is larger than that for case with $d_{D,0}=0.1$ for the same $\beta_0$, i.e. the period for each cycle is shorter for smaller $d_{D,0}$ case. On the other hand, as previously discussed, the growth rate for $d_{D,0}=0.01$ case is smaller than that for $d_{D,0}=0.1$ case. As a consequence, these two aspects lead to a larger extent of suppression for $d_{D,0}=0.01$ case under the same magnetic field (see Figs. \ref{fig:gr_dd0_1_mag}(b) and \ref{fig:gr_dd0_01_mag}(b) ). The same process occurs in the electrons though not elaborated here. The perturbation amplitude of both the ion and electrons (see Fig. \ref{fig:gr_dd0_01_mag}(b) and (d)) are virtually identical (except for the very short duration just after the ion shock interaction) due to the strong coupling between the charged species in this case.  

\section{Summary and Conclusion\label{sec:5}}
In this work, we investigate the linear evolution of RM instability in the framework of an ideal two-fluid plasma model. By separating the original equations into base and perturbation parts, we first compute the nonlinear base state, then solve the linearized equations governing the perturbed state. The base state Lorentz force is an important forcing term in the dynamics of the perturbations. The non-dimensional Debye length $d_{D,0}$ governs the level of coupling between the ions and electrons. 
By varying the reference Debye length $d_{D,0}$, we examine the two-fluid effect on the RM instability that occurs when an ion shock interacts with the ion density interface. 
When $d_{D,0}$ is large, the coupling between ions and electrons is sufficiently small that the induced Lorentz force is too weak to influence the particles. In this scenario, the two species evolve as two separate fluids. When $d_{D,0}$ is small, the coupling is strong and the induced Lorentz force is strong enough that the difference between state of ions and electrons is rapidly decreased by the force. As a consequence, the ions and electrons are tightly coupled and evolve like one fluid. The evolution of growth rate and amplitude of interfaces is investigated for the cases with different $d_{D,0}$. 
Temporally, we distinguish between an early phase during which electron precursor waves interact with the electron interface, and the second instability phase when the ion shock interacts with the ion density interface causing the perturbation to grow. The electron precursor waves induce an oscillating force on the perturbations and the growth rate oscillates about a zero mean. 
%We show that the interaction between electron waves and interface induces an oscillating force acting on perturbations, results in the growth rate oscillating around $0$. Meanwhile, the perturbation amplitude of ions and electrons oscillates around $0$ with a very small amplitude. 
After the ion shock-interface interaction, the growth rate and amplitude of perturbations almost ``linearly" increase. When $d_{D,0}$ is small, the induced force acting on the perturbations is strong while the duration this force acts shortens.  This forms a competitive mechanism in the development of growth rate. In the case with $d_{D,0}=0.01$, the latter dominates so that the final growth rate of each species is less than those of the case with $d_{D,0}=0.1$. 
We also examine the effect of an initially applied magnetic field in the streamwise direction characterized by the non-dimensional parameter $\beta_0$. The magnetic field has a very small influence on the perturbations during the electron precursor waves interaction with the interface. For a short duration after the ion shock interaction, the growth rate is very similar for different initial magnetic field strengths. However, as time progresses, the suppression of the instability is observed. Moreover, the time duration taken for the instability to be suppressed is directly proportional to $\beta_0$. The growth rate shows oscillations with a frequency that is related to the ion or electron cyclotron frequency. 
 For the smaller value of  $d_{D,0}=0.01 $ the ion and electron perturbation amplitude history are virtually identical.  Both the growth rate and amplitude of perturbations are suppressed due to the vorticity on interfaces is transported away from the interface.
 %%%%%%%%%%%%%%%%%%%%%%%%%%%%%%%%%%%%%%%%%%%%%%%%%%%%%%%%%%%%%%%%%%%%%%
\begin{acknowledgments}
This research was supported by the KAUST Office of Sponsored Research under Award
URF/1/3418-01.
\end{acknowledgments}
\appendix\section{\label{apdx_matrix}Matrices $A$, $B$ and $C$}

The matrices in {Eq.}~(\ref{eq:streqpert}) which arise after the linearization of the governing equations are presented in their full form below. 
$A=
\begin{pmatrix}
 A_i & 0 & 0  \\
 0 & A_e & 0  \\
 0 & 0 & A_{EM} \\
\end{pmatrix}
$, 
$A_{EM}=
\begin{pmatrix}
 0 & 0 & 0 & 0 & 0 & 0 \\
 0 & 0 & 0 & 0 & 0 & -c \\
 0 & 0 & 0 & 0 & c & 0 \\
 0 & 0 & 0 & 0 & 0 & 0 \\
 0 & 0 & c & 0 & 0 & 0 \\
 0 & -c & 0 & 0 & 0 & 0 \\
\end{pmatrix}
$\\
\begin{frame}

\resizebox{\linewidth}{!}{%
$A_\alpha=
\begin{pmatrix}
 0 & 1 & 0 & 0 & 0 \\
 \frac{1}{2} \left((\gamma -3) u_\alpha^2+(\gamma -1) \left(v_\alpha^2+w_\alpha^2\right)\right) & -(\gamma -3) u_\alpha & -(\gamma -1) v_\alpha & -(\gamma -1) w_\alpha & \gamma -1  \\
 -u_\alpha v_\alpha & v_\alpha & u_\alpha & 0 & 0  \\
 -u_\alpha w_\alpha & w_\alpha & 0 & u_\alpha & 0  \\
 \frac{u_\alpha \left(\left(\gamma ^2-3 \gamma +2\right) \left(u_\alpha^2+v_\alpha^2+w_\alpha^2\right) \rho _\alpha-2 \gamma  p_\alpha\right)}{2 (\gamma -1) \rho _\alpha} & \frac{2 \gamma  p_\alpha-(\gamma -1) \left((2 \gamma -3) u_\alpha^2-v_\alpha^2-w_\alpha^2\right) \rho _\alpha}{2 (\gamma -1) \rho _\alpha} & -(\gamma -1) u_\alpha v_\alpha & -(\gamma -1) u_\alpha w_\alpha & \gamma  u_\alpha \\
\end{pmatrix}
$}
\end{frame}\\

$B=
\begin{pmatrix}
 B_i & 0 & 0  \\
 0 & B_e & 0  \\
 0 & 0 & B_{EM} \\
\end{pmatrix}
$,
$B_{EM}=
\begin{pmatrix}

 0 & 0 & 0 & 0 & 0 & c \\
 0 & 0 & 0 & 0 & 0 & 0 \\
 0 & 0 & 0 & -c & 0 & 0 \\
 0 & 0 & -c & 0 & 0 & 0 \\
 0 & 0 & 0 & 0 & 0 & 0 \\
 c & 0 & 0 & 0 & 0 & 0 \\

\end{pmatrix}
$

\begin{frame}

\resizebox{\linewidth}{!}{%
$B_\alpha=
\begin{pmatrix}

 0 & 0 & 1 & 0 & 0 \\
 -u_\alpha v_\alpha & v_\alpha & u_\alpha & 0 & 0 \\
 \frac{1}{2} \left((\gamma -1) u_\alpha^2+(\gamma -3) v_\alpha^2+(\gamma -1) w_\alpha^2\right) & -(\gamma -1) u_\alpha & -(\gamma -3) v_\alpha & -(\gamma -1) w_\alpha & \gamma -1 \\
 -v_\alpha w_\alpha & 0 & w_\alpha & v_\alpha & 0 \\
 \frac{v_\alpha \left(\left(\gamma ^2-3 \gamma +2\right) \left(u_\alpha^2+v_\alpha^2+w_\alpha^2\right) \rho _\alpha-2 \gamma  p_\alpha\right)}{2 (\gamma -1) \rho _\alpha} & -(\gamma -1) u_\alpha v_\alpha & \frac{2 \gamma  p_\alpha+(\gamma -1) \left(u_\alpha^2+(3-2 \gamma ) v_\alpha^2+w_\alpha^2\right) \rho _\alpha}{2 (\gamma -1) \rho _\alpha} & -(\gamma -1) v_\alpha w_\alpha & \gamma  v_\alpha \\

\end{pmatrix}
$
}
\end{frame}\\

\begin{frame}

\resizebox{\linewidth}{!}{%
$C= \\ \\
\left(
\begin{array}{cccccccccccccccc}
 0 & 0 & 0 & 0 & 0 & 0 & 0 & 0 & 0 & 0 & 0 & 0 & 0 & 0 & 0 & 0 \\ 
 \frac{q_i E_x}{c d_{D,0} m_i} & 0 & \frac{q_i B_z}{c d_{D,0} m_i} & -\frac{q_i B_y}{c d_{D,0} m_i} & 0 & 0 & 0 & 0 & 0 & 0 & 0 & -\frac{q_i w_i \rho _i}{c d_{D,0} m_i} & \frac{q_i v_i \rho _i}{c d_{D,0} m_i} & \frac{q_i \rho _i}{c d_{D,0} m_i} & 0 & 0 \\ 
 \frac{q_i E_y}{c d_{D,0} m_i} & -\frac{q_i B_z}{c d_{D,0} m_i} & 0 & \frac{q_i B_x}{c d_{D,0} m_i} & 0 & 0 & 0 & 0 & 0 & 0 & \frac{q_i w_i \rho _i}{c d_{D,0} m_i} & 0 & -\frac{q_i u_i \rho _i}{c d_{D,0} m_i} & 0 & \frac{q_i \rho _i}{c d_{D,0} m_i} & 0 \\ 
 \frac{q_i E_z}{c d_{D,0} m_i} & \frac{q_i B_y}{c d_{D,0} m_i} & -\frac{q_i B_x}{c d_{D,0} m_i} & 0 & 0 & 0 & 0 & 0 & 0 & 0 & -\frac{q_i v_i \rho _i}{c d_{D,0} m_i} & \frac{q_i u_i \rho _i}{c d_{D,0} m_i} & 0 & 0 & 0 & \frac{q_i \rho _i}{c d_{D,0} m_i} \\ 
 0 & \frac{q_i E_x}{c d_{D,0} m_i} & \frac{q_i E_y}{c d_{D,0} m_i} & \frac{q_i E_z}{c d_{D,0} m_i} & 0 & 0 & 0 & 0 & 0 & 0 & 0 & 0 & 0 & \frac{q_i u_i \rho _i}{c d_{D,0} m_i} & \frac{q_i v_i \rho _i}{c d_{D,0} m_i} & \frac{q_i w_i \rho _i}{c d_{D,0} m_i} \\ 
 0 & 0 & 0 & 0 & 0 & 0 & 0 & 0 & 0 & 0 & 0 & 0 & 0 & 0 & 0 & 0 \\ 
 0 & 0 & 0 & 0 & 0 & \frac{q_e E_x}{c d_{D,0} m_e} & 0 & \frac{q_e B_z}{c d_{D,0} m_e} & -\frac{q_e B_y}{c d_{D,0} m_e} & 0 & 0 & -\frac{q_e w_e \rho _e}{c d_{D,0} m_e} & \frac{q_e v_e \rho _e}{c d_{D,0} m_e} & \frac{q_e \rho _e}{c d_{D,0} m_e} & 0 & 0 \\ 
 0 & 0 & 0 & 0 & 0 & \frac{q_e E_y}{c d_{D,0} m_e} & -\frac{q_e B_z}{c d_{D,0} m_e} & 0 & \frac{q_e B_x}{c d_{D,0} m_e} & 0 & \frac{q_e w_e \rho _e}{c d_{D,0} m_e} & 0 & -\frac{q_e u_e \rho _e}{c d_{D,0} m_e} & 0 & \frac{q_e \rho _e}{c d_{D,0} m_e} & 0 \\ 
 0 & 0 & 0 & 0 & 0 & \frac{q_e E_z}{c d_{D,0} m_e} & \frac{q_e B_y}{c d_{D,0} m_e} & -\frac{q_e B_x}{c d_{D,0} m_e} & 0 & 0 & -\frac{q_e v_e \rho _e}{c d_{D,0} m_e} & \frac{q_e u_e \rho _e}{c d_{D,0} m_e} & 0 & 0 & 0 & \frac{q_e \rho _e}{c d_{D,0} m_e} \\
 0 & 0 & 0 & 0 & 0 & 0 & \frac{q_e E_x}{c d_{D,0} m_e} & \frac{q_e E_y}{c d_{D,0} m_e} & \frac{q_e E_z}{c d_{D,0} m_e} & 0 & 0 & 0 & 0 & \frac{q_e u_e \rho _e}{c d_{D,0} m_e} & \frac{q_e v_e \rho _e}{c d_{D,0} m_e} & \frac{q_e w_e \rho _e}{c d_{D,0} m_e} \\ 
 0 & 0 & 0 & 0 & 0 & 0 & 0 & 0 & 0 & 0 & 0 & 0 & 0 & 0 & 0 & 0 \\ 
 0 & 0 & 0 & 0 & 0 & 0 & 0 & 0 & 0 & 0 & 0 & 0 & 0 & 0 & 0 & 0 \\ 
 0 & 0 & 0 & 0 & 0 & 0 & 0 & 0 & 0 & 0 & 0 & 0 & 0 & 0 & 0 & 0 \\ 
 0 & -\frac{c q_i}{d_{D,0} m_i} & 0 & 0 & 0 & 0 & -\frac{c q_e}{d_{D,0} m_e} & 0 & 0 & 0 & 0 & 0 & 0 & 0 & 0 & 0 \\ 
 0 & 0 & -\frac{c q_i}{d_{D,0} m_i} & 0 & 0 & 0 & 0 & -\frac{c q_e}{d_{D,0} m_e} & 0 & 0 & 0 & 0 & 0 & 0 & 0 & 0 \\ 
 0 & 0 & 0 & -\frac{c q_i}{d_{D,0} m_i} & 0 & 0 & 0 & 0 & -\frac{c q_e}{d_{D,0} m_e} & 0 & 0 & 0 & 0 & 0 & 0 & 0 \\ 
 \end{array}
\right)$}
\end{frame}

{{
\section{\label{apdx_conv}Convergence test}
\begin{figure}%[!ht]
\centering
\includegraphics[width=\linewidth]{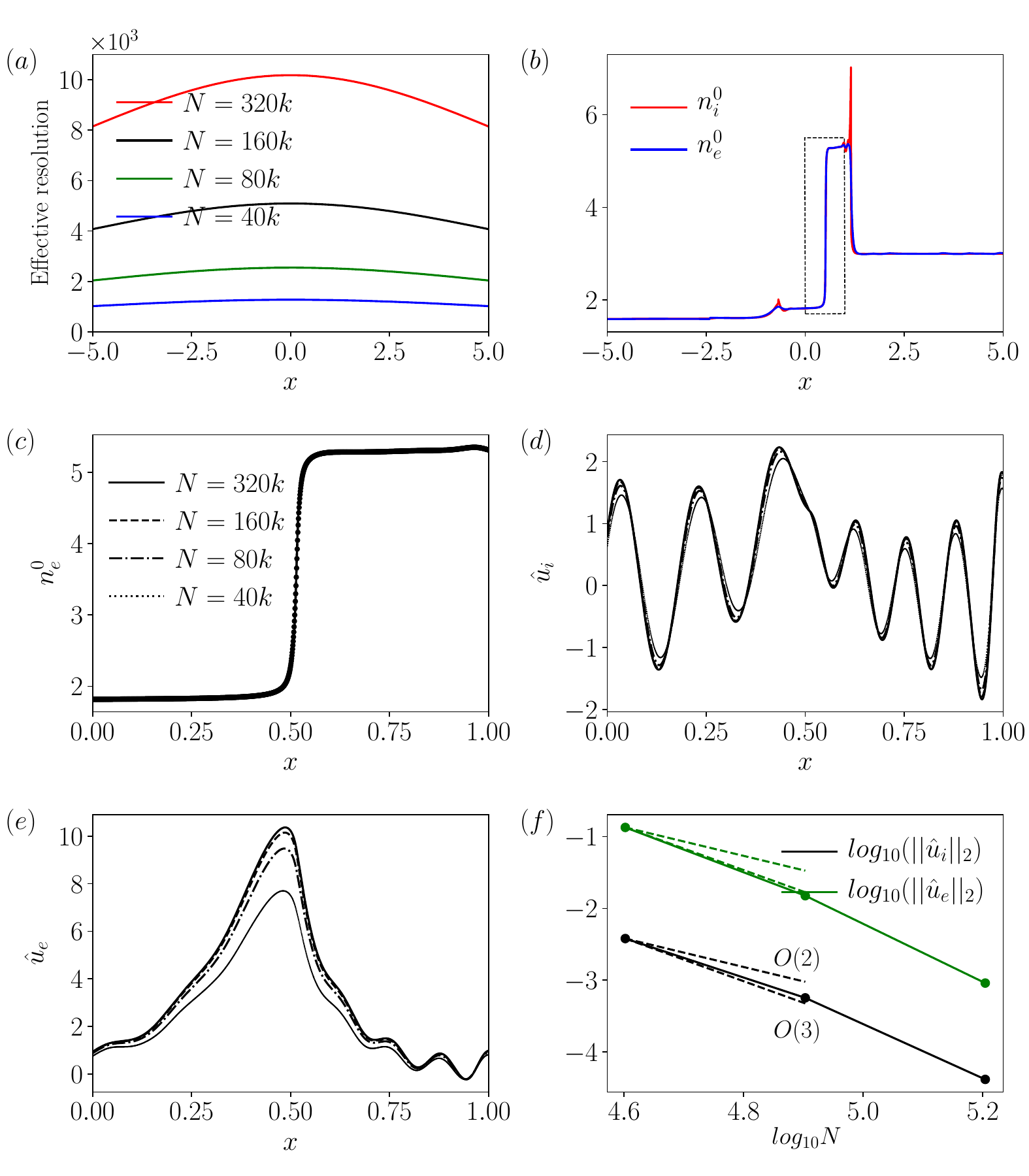}
\caption{\label{figure_converg1d} (a)Effective resolution for various mesh size. (b) Ion and electron base number density at $t=3.77$ from simulation. Zoom in of detail showing (c) electron base number density $n_e^0$, (d) perturbed ion $x-$ velocity $\hat{u}_i$, and (e) perturbed electron $x-$ velocity with increasing mesh size. (f) $L^2$ norm error of perturbed velocities.}
\end{figure}
Figure~\ref{figure_converg1d}(a) plots the effective resolution of the linear two-fluid plasma cases with various mesh sizes. Both the ion and electron density interfaces have not travelled across the position $x=5$ at the end of the simulation time.  Therefore we only plot the effective resolution in the domain $x\in (-5,5)$. We can see that when the mesh size $N=160000$, the effective resolution is at least $4000$ per unit length during our simulations. At $t=3.77$, the ion shock has interacted with the ion interface, as shown in the fig.~\ref{figure_converg1d}(b). To demonstrate grid convergence, the simulation results ( $n_e^0$, $\hat{u}_i$ and $\hat{u}_e$) in the region near density interface ($x\in(0,1)$) are considered, as shown in the figs.~\ref{figure_converg1d}(c), (d) and (e). The $L_2$ norm of the difference from the finest grid solution suggests a convergence rate of order at least $3$ (see fig.~\ref{figure_converg1d}(f)). It shows that a mesh size of  $N=160000$ captures the essential details of the flow. Thus the minimum effective resolution adopted in our simulations is $4000$.

\section{\label{apdx_comparison}Comparison between linear and nonlinear simulations}
In this section, a comparison between linear and nonlinear simulations reported in paper by Bond {\it et al} \citep{bond2020} will be conducted by considering  the ion vorticity on the interface. The perturbed ion vorticity evolution equation is given as (here we apply the dimensionless parameters as in paper \citep{bond2017} for convenience), 
\begin{equation}
\dfrac{D\hat{\bm{\omega}}_i}{Dt} =\underbrace{ \widehat{(\bm{\omega}_i\cdot\nabla)\bm{u}_i}}_{\hat{\bm{\tau}}_{v,i}}+ \underbrace{\widehat{\bm{\omega}_i(-\nabla\cdot\bm{u}_i)}}_{\hat{\bm{\tau}}_{s,i}}+\underbrace{\widehat{\frac{\nabla\rho_i \times \nabla p_i}{\rho_i^2} }}_{\hat{\bm{\tau}}_{b,i}}+\underbrace{\dfrac{r_i c}{d_{L,0}}\widehat{\nabla\times\bm{E}}}_{\hat{\bm{\tau}}_{E,i}}+\underbrace{\dfrac{r_i}{d_{L,0}}\widehat{\nabla\times(\bm{u}_i\times\bm{B})}}_{\hat{\bm{\tau}}_{B,i}}
\end{equation}
where $r_i = \frac{q_i}{m_i}$ and $d_{L,0}$ is the reference Larmor radius. From the equation, the contribution of the perturbed vorticity can be split into five parts,

\noindent $\hat{\bm{\tau}}_{v,i}=(\hat{\bm{\omega}}_i\cdot\nabla)\bm{u}^0_i$: vorticity stretching term,

\noindent $\hat{\bm{\tau}}_{s,i}=-\hat{\bm{\omega}}_i(\nabla\cdot\bm{u}^0_i)$: compressibility effects on vorticity,

\noindent $\hat{\bm{\tau}}_{b,i}=\frac{1}{(\rho_i^0)^2}(\nabla \rho_i^0\times\nabla\hat{p}_i+\nabla \hat{\rho}_i\times\nabla p_i^0)$: baroclinic torque,

\noindent $\hat{\bm{\tau}}_{E,i}=\frac{r_i c}{d_{L,0}}\nabla\times\hat{\bm{E}}$: torque from electric contributions to the Lorentz force,

\noindent $\hat{\bm{\tau}}_{B,i}=\frac{r_i}{d_{L,0}}\nabla\times(\bm{u}_i^0\times\hat{\bm{B}}+\hat{\bm{u}}_i\times\bm{B}^0)$: torque from magnetic contributions to the Lorentz force.

\begin{figure}%[!ht]
\centering
\includegraphics[width=\linewidth]{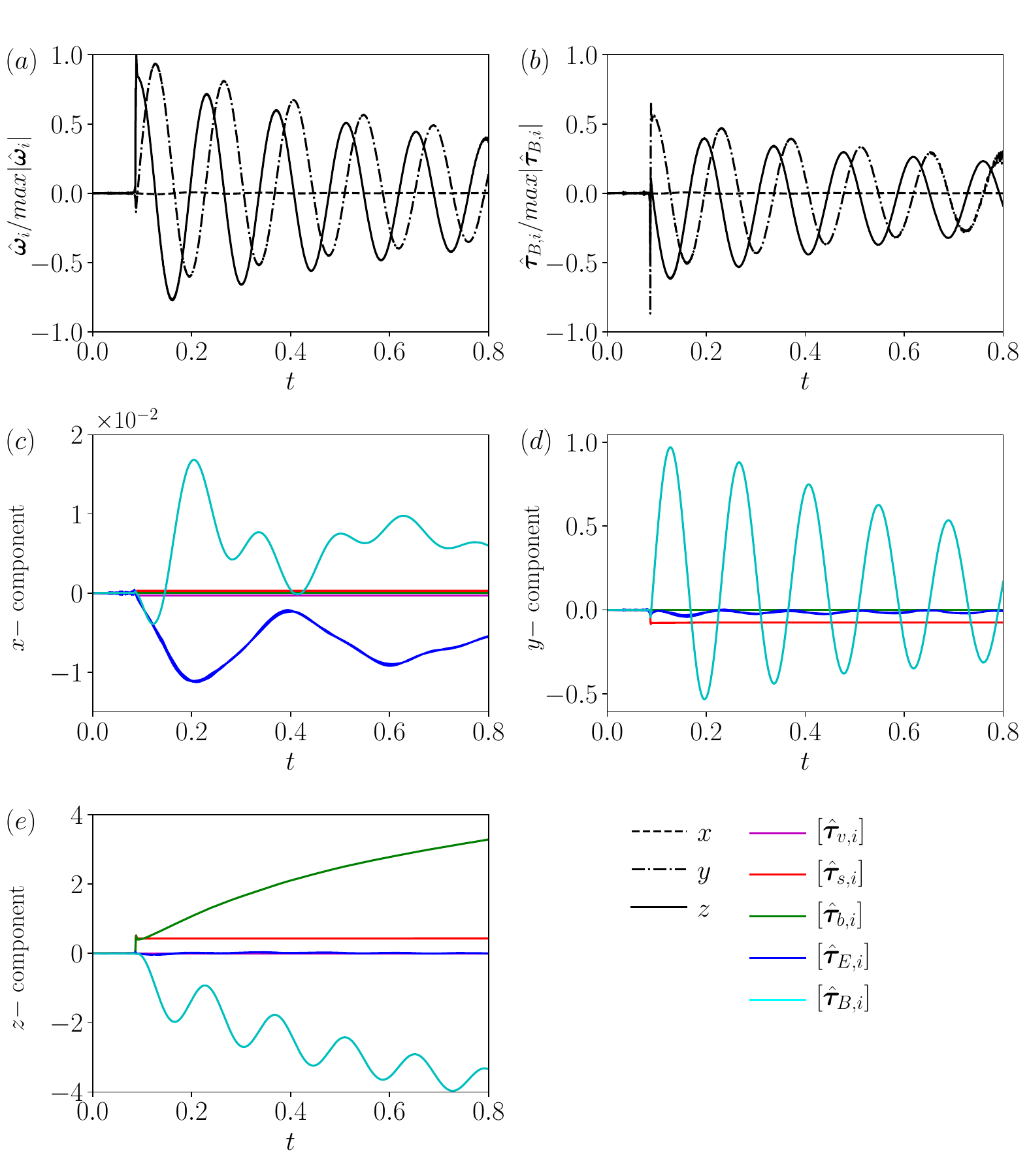}
\caption{\label{figure_vortauB} Evolution of (a) perturbed ion vorticity $\hat{\bm{\omega}}_i$ and (b) perturbed ion torque $\hat{\bm{\tau}}_{B,i}$ due to the magnetic field for the case with reference Debye length $d_{D,0}=0.002$, reference Larmor radius $d_{L,0}=0.0707$ and $\beta_0=0.1$ (equivalent initial conditions as the nonlinear case in the Fig. 12,  paper by Bond {\it et al} \citep{bond2020}). The contributions of each source term are compared in (c) x, (d) y, and (e) z directions. The  `$[~]$' operator is defined as $[ \xi ] := \int_0^t \xi d \tau /max|\hat{\bm{\omega}}_i|$. }
\end{figure}

Figure~\ref{figure_vortauB}(a) plots the evolution of the three components of perturbed vorticity $\hat{\bm{\omega}}_i$ on the ion density interface for the case with same initial conditions as the nonlinear case in the Fig. 12,  paper by Bond {\it et al} \citep{bond2020}.
It shows that the baroclinic torque $\hat{\bm{\tau}}_{b,i}$ which has a contribution only in the $z$ direction, is one main source for the perturbed ion vorticity on the ion interface during the ion shock-interface interaction. Meanwhile, the contribution of vorticity due to compressibility $\hat{\bm{\tau}}_{s,i}$ generates vorticity and deposits on the interface mainly in $z$ direction along with smaller contributions in $x-y$ plane. As a result, $\hat{\omega}_{z,i}$ rapidly grows while the other two components remain small at $t\approx 0.1$, as shown in Figs.~\ref{figure_vortauB}(a, c, d, e).
After the interaction, the perturbed $y-$ velocity $\hat{v}_i$ induced by the large deposited $z-$ vorticity $\hat{\omega}_{z,i}$ interacts with the strong base magnetic field $B_x^0$, resulting in the perturbed Lorentz force in $z$ direction that induces the perturbed $z-$ velocity $\hat{w}_i$. Following the similar process, the resulting $\hat{w}_i$ induces $\hat{v}_i$ in return. 
As a result, $\hat{v}_i$ and $\hat{w}_i$ oscillates in time due to the above circular motion and a relative phase angle occurs in $y-z$ plane, same as for  $\hat{\bm{\omega}}_i$ and $\hat{\bm{\tau}}_{b,i}$ (see Figs.~\ref{figure_vortauB} (a, b)). 
Due to the existence of $B_x^0$ and $u^0$ only in the base state, the above dynamics take place mainly in $y-z$ plane, with much smaller contribution in the $x$ direction.

On account of the lack of nonlinear effects in our simulations, it is obvious that the evolution of $\hat{\bm{\omega}}_i$ in the linear case is not same as that in nonlinear case, especially for the $x-$ component. A comparison between linear and nonlinear simulations is qualitatively addressed by considering the following. For instance, in nonlinear simulation, both vorticity and torque lie most in the $y-z$ plane and oscillate over time with decreasing in magnitude, which can also be observed in the linear case. Moreover, by $t=0.8$, there are about $5.1$ and $5$ cycles along $y$ and $z$ directions for $\hat{\bm{\omega}}_i$, respectively (about $4.7$ and $4.7$ in Bond's results) while about $5$ and $5.1$ cycles respectively for $\hat{\bm{\tau}}_{B,i}$ (about $4.7$ and $4.7$ in Bond's results), which are consistent with nonlinear results. Thus the frequency on $\hat{\bm{\omega}}_i$ ($y$ and $z$ components) matches well between the linear and nonlinear simulations.
}}
\nocite{*}
\bibliography{references}% Produces the bibliography via BibTeX.

\end{document}